\documentclass[sigconf]{acmart}
\newcommand{\system}{\textsc{InTune}}
\AtBeginDocument{%
  }
    
\usepackage{amsmath}
\usepackage{tabularx}

\newcommand{\code}[1]{\texttt{#1}}
\copyrightyear{2023}
\acmYear{2023}
\setcopyright{rightsretained}
\acmConference[RecSys '23]{Seventeenth ACM Conference on Recommender Systems}{September 18--22, 2023}{Singapore, Singapore}
\acmBooktitle{Seventeenth ACM Conference on Recommender Systems (RecSys '23), September 18--22, 2023, Singapore, Singapore}\acmDOI{10.1145/3604915.3608778}
\acmISBN{979-8-4007-0241-9/23/09}

\usepackage{array}
\usepackage{listings}
\usepackage{pifont}% http://ctan.org/pkg/pifont
\usepackage{subfigure,xspace,verbatim}
\usepackage{hyperref,color,balance,multirow}
\usepackage{balance,float,url,amsfonts,alltt}
\newcolumntype{P}[1]{>{\centering\arraybackslash}p{#1}}
\definecolor{darkgreen}{RGB}{6,92,9}

\definecolor{codegreen}{rgb}{0,0.6,0}
\definecolor{codegray}{rgb}{0.5,0.5,0.5}
\definecolor{codepurple}{rgb}{0.58,0,0.82}
\definecolor{backcolour}{rgb}{1.0,1.00,1.00}

\lstdefinestyle{mystyle}{
    backgroundcolor=\color{backcolour},   
    commentstyle=\color{codegreen},
    keywordstyle=\color{magenta},
    numberstyle=\tiny\color{codegray},
    stringstyle=\color{codepurple},
    basicstyle=\ttfamily\tiny,
    breakatwhitespace=false,         
    breaklines=true,                 
    captionpos=b,                    
    keepspaces=true,                 
    numbers=left,                    
    numbersep=5pt,                  
    showspaces=false,                
    showstringspaces=false,
    showtabs=false,                  
    tabsize=2
}

\begin{document}

\title{\system: Reinforcement Learning-based Data Pipeline Optimization for Deep Recommendation Models}

\author{Kabir Nagrecha}
\email{knagrecha@netflix.com}
\affiliation{%
  \institution{Netflix, Inc.}
  \city{Los Gatos, CA}
  \country{USA}
}

\author{Lingyi Liu}
\email{lliu@netflix.com}
\affiliation{%
  \institution{Netflix, Inc.}
  \city{Los Gatos, CA}
  \country{USA}
}

\author{Pablo Delgado}
\email{pdelgado@netflix.com}
\affiliation{%
  \institution{Netflix, Inc.}
  \city{Los Gatos, CA}
  \country{USA}
}

\author{Prasanna Padmanabhan}
\email{ppadmanabhan@netflix.com}
\affiliation{%
  \institution{Netflix, Inc.}
  \city{Los Gatos, CA}
  \country{USA}
}
\renewcommand{\shortauthors}{Nagrecha et al.}

\begin{abstract}
Deep learning-based recommender models (DLRMs) have become an essential component of many modern recommender systems.
Several companies are now building large compute clusters reserved only for DLRM training, driving new interest in cost- \& time- saving optimizations.
The systems challenges faced in this setting are unique; while typical deep learning (DL) training jobs are dominated by model execution times, the most important factor in DLRM training performance is often \textit{online data ingestion.}

In this paper, we explore the unique characteristics of this data ingestion problem and provide insights into the specific bottlenecks and challenges of the DLRM training pipeline at scale. 
We study real-world DLRM data processing pipelines taken from our compute cluster at Netflix to both observe the performance impacts of online ingestion and to identify shortfalls in existing data pipeline optimizers.
We find that current tooling either yields sub-optimal performance, frequent crashes, or else requires impractical cluster re-organization to adopt.
Our studies lead us to design and build a new solution for data pipeline optimization, \system.
\system~employs a reinforcement learning (RL) agent to learn how to distribute the CPU resources of a trainer machine across a DLRM data pipeline to more effectively parallelize data-loading and improve throughput.
Our experiments show that \system~can build an optimized data pipeline configuration within only a few minutes, and can easily be integrated into existing training workflows.
By exploiting the responsiveness and adaptability of RL, \system~achieves significantly higher online data ingestion rates than existing optimizers, thus reducing idle times in model execution and increasing efficiency.
We apply \system~to our real-world cluster, and find that it increases data ingestion throughput by as much as 2.29X versus current state-of-the-art data pipeline optimizers while also improving both CPU \& GPU utilization.
\end{abstract}



\maketitle

\section{Introduction}\label{sec:intro}
Recommendation systems now underpin many essential components of the web ecosystem, including search result ranking, e-commerce product placement, and media suggestions in streaming services.
Over the last several years, many of these services have begun to employ \textit{deep learning} (DL) models in their recommendation infrastructure, to better exploit historical patterns in their data.
In turn, DL-based product recommendation has quickly become one of the most commercially significant applications of DL.
Companies have begun to invest heavily in DL recommendation infrastructure, often maintaining entire datacenters and super-clusters for the sole purpose of recommender model training~\cite{acun2021understanding}.
But in many cases, these infrastructural investments have run into critical hurdles~\cite{gupta2020architectural}.
Practitioners and cluster administrators are discovering that the \textit{training optimization} challenges faced with DL recommender models differ significantly from those seen in historical practice with other DL model types~\cite{wu2021sustainable}.
In particular, recent studies of industry clusters have found that the unique design of recommender model architectures has left training pipelines susceptible to inefficiencies in \textit{data ingestion}~\cite{Zhao_2022}.

Most DL architectures are dominated by high-intensity matrix operators, and standard tooling for DL training optimization has evolved to support models that fit this pattern~\cite{megatron2019,pipedream2018,flexflow2018,gshard2020,zerooffload2021,gpipe2018,researchExam}.
In such cases, model execution usually dominates training times to such a degree that data ingestion procedures (e.g. disk loading, shuffling, etc) can be overlapped with and hidden underneath the matrix operation times.
Unfortunately, however, DL-based recommender models (DLRMs) are atypical in this regard.

\begin{figure}[h]
\includegraphics[width=\columnwidth]{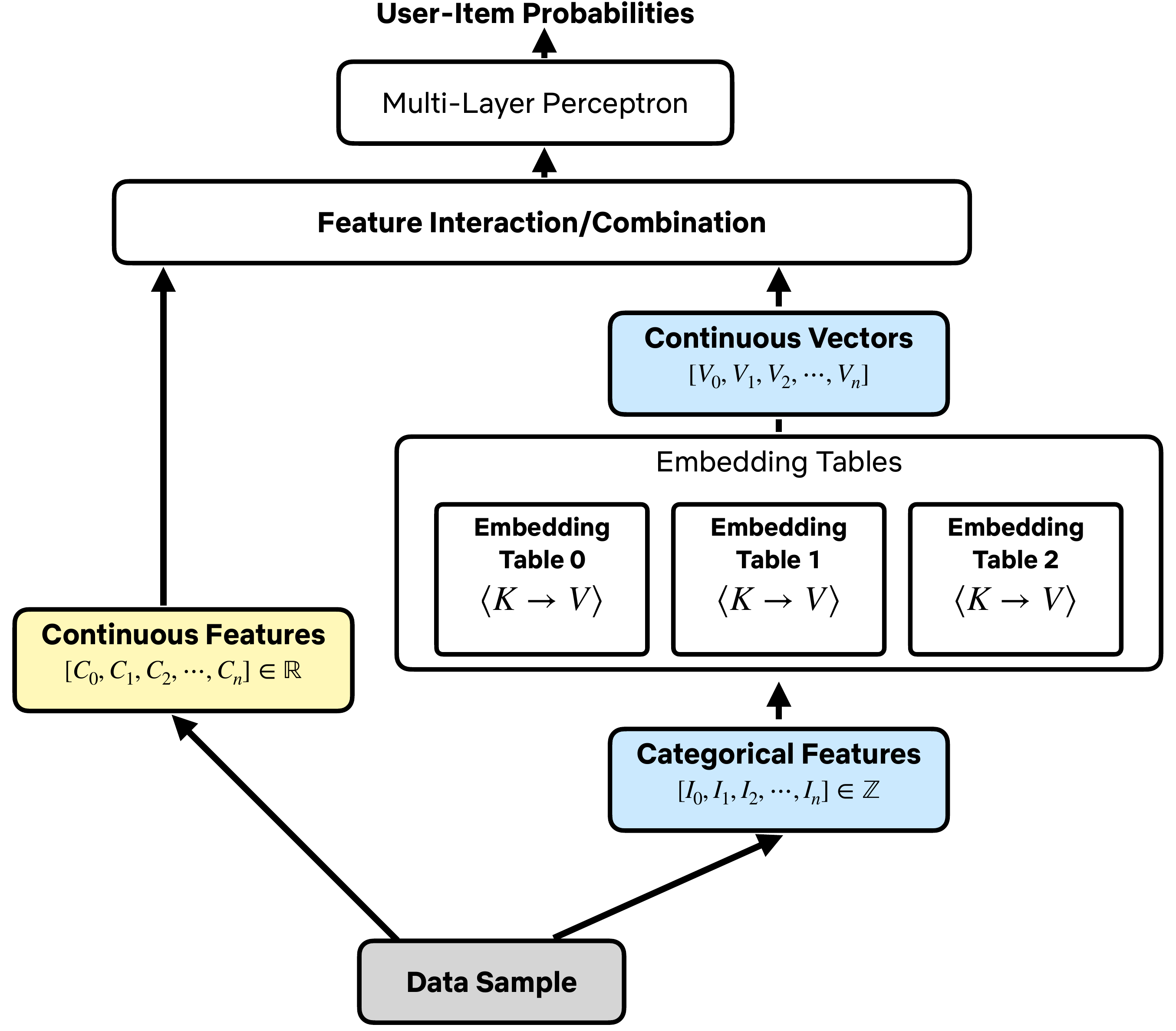}
\caption{A typical DLRM architecture~\cite{cheng2016wide,acun2021understanding}. The model uses an embedding table to convert sparse categorical data to dense vectors that can then be merged with dense features in some overlaid DNN. Adapted from a similar illustration in prior art~\cite{Wei_2022}.}\label{fig:dlrm_arch_distribution}
\Description[The DLRM architecture has a table in it.]{Illustration of a DLRM architecture shows that the single most important component is the embedding table.}
\end{figure}

Recommender datasets are generally composed of both sparse (categorical) and dense (continuous) features, and joining information across features requires transforming these two representations into a common format.
To this end, DLRM architectures use \textit{embedding tables} to transform categorical inputs into dense embedding vectors through a hash-table lookup.
These can then be combined with the dense vectors and fed through some secondary DL model to produce user-item probability ratings~\cite{Wei_2022}.
Figure~\ref{fig:dlrm_arch_distribution} illustrates a typical architecture.

\begin{figure*}[h]
\includegraphics[width=\textwidth]{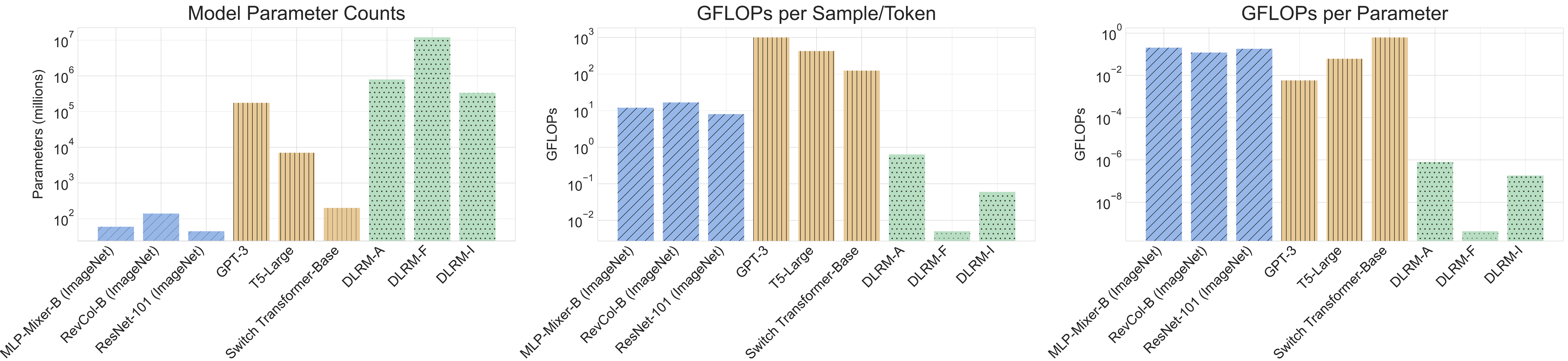}
\caption{Approximate parameter \& FLOP counts for popular architectures in language modeling and image recognition contrasted against DLRM models drawn from a recent paper~\cite{mudigere2021software}. FLOPs are reported on single-element batches (single-token for language models).  We also report averaged FLOPs per parameter, derived from the previous two charts. Y-axis is set to log-scale for all charts.}\label{fig:flops}
\Description[DLRMs are much less computationally intensive than other architectures.]{Comparison of FLOPs \& parameter counts across model architecture types shows that DLRMs are far less computationally intensive for their size than other architectures.}
\vspace{-5mm}
\end{figure*}

%\vspace{1mm}
The embedding tables, which are the typically single largest component of the DLRM architecture, use a key-value lookup rather than dense matrix multiplication.
For this reason, DLRM models are often less compute-intensive than other architectures of a comparable size.
Figure~\ref{fig:flops} charts the differences in computational intensity between large-scale recommendation models versus language models and computer vision architectures, illustrating that DLRM models require \textit{orders of magnitude} fewer operations than comparably-sized Transformers or Convolutional Neural Networks. This uniquely light computational footprint can lead to unexpected system optimization challenges.

\vspace{3mm}

\textbf{Challenge \& Motivation.} The low computational intensity of DLRMs generally translates to low model latencies, which fail to mask the cost of data-loading and transformation.
Improved GPU hardware and new model acceleration techniques have only exacerbated this issue by reducing model runtimes and increasing the requisite data-loading throughput to keep the model fed during training.
The fault is not only with the models, however; the problem is aggravated further still by the generally high demands of \textit{online data processing}, i.e. data transformation at ingestion time, for recommendation applications.
In other domains (e.g. language modeling, computer vision) not only are training times dominated by model execution, so that data processing latencies can be more effectively hidden, but it is also practical to push the heaviest data transformation steps to an offline pre-compute phase. This greatly reduces the need to optimize data-loading. By contrast, recommendation datasets are uniquely reliant on the \textit{online} step, which must be done alongside model execution. We attribute this to three characteristics of recommendation data: \textbf{scale}, \textbf{reusability}, and \textbf{volatility}.

\vspace{1mm}

First, \textit{scale}. A recommender dataset for a popular application might span billions of interactions and require terabytes (or even petabytes!) of disk space.
Offline data transformation can bloat these already high storage costs further still. 
Consider, for example, a common data processing operation such as \textit{augmentation}, which randomly modifies various aspects of a data sample to produce an altogether new sample. 
Applying this operation offline might double or triple the size of an already massive dataset; the only practical way to run such transformations would be to do them online so that augmented samples can be discarded as soon as they are consumed.
Furthermore, this scale issue often makes \textit{caching}, which might otherwise help to mitigate processing challenges, impractical.

Second, \textit{reusability}. A single core dataset might be reused for multiple different DLRM architectures. In a movie recommendation system, one DLRM might be used to rank rows, another might be used to rank search results, while yet another might rank genres. Each model would likely require different data transformation operations and feature extraction procedures. Pushing data transformation to the offline phase would require replicating and re-processing the original dataset dozens of times, again bloating storage and compute costs.

Third, \textit{volatility}. Recommendation datasets are updated frequently as new interactions are recorded. In addition, \textit{ephemeral IDs} often lead to dataset changes in domains such as e-commerce, e.g. when a product is added or removed from the platform. Any offline transformations would have to be re-run frequently as the dataset evolves. Incremental transformation is not always practical; some operations such as shuffling require the whole dataset to be present.

\vspace{1mm}

Prior analyses~\cite{Zhao_2022,mohan2020analyzing} of DLRM training have recorded the impacts of these issues in practice, suggesting that online data ingestion optimization is critical to improving DLRM training performance. This new and emerging problem lacks a satisfactory solution. Table~\ref{tb:prior_art} provides an overview of existing tooling, but none of these prior systems can effectively tackle this issue. Generic pipeline tools such as \code{AUTOTUNE} \& Plumber~\cite{murray2021tf,kuchnik2022plumber} often lead to sub-optimal performance for DLRM jobs, or can even cause fatal out-of-memory errors. GPU data-loaders can be situationally useful, but cannot be recommended for general use due to concerns over processor cycle contention between the model and pipeline~\cite{Zhao_2022}. The only CPU-based DLRM data pipeline work we are aware of~\cite{Zhao_2022,mudigere2021software} relies on a specialized cluster architecture design, and is not feasible to adopt for typical users. We expand on these in Section~\ref{sec:prior_art}.

\begin{table*}[t]
\centering
\caption{Overview of existing tooling.}
\vspace{-1mm}
\begin{tabularx}{\textwidth}{ X P{2.5cm} P{9cm} } 
& Name & Description  \\
 \toprule 
\multirow{3}{*}{\textbf{\shortstack{Generic Pipelines}}} & \code{AUTOTUNE}~\cite{murray2021tf} & TensorFlow's built-in tool for optimizing \code{tf.data} pipelines, considered to be a state-of-the-art optimizer~\cite{kuchnik2022plumber}.  \\ 
 							       & Plumber~\cite{kuchnik2022plumber} & \code{AUTOTUNE} alternative with roughly equivalent performance.\\ 
                                                                                \midrule
\textbf{\shortstack{DLRM Pipelines}} & Data PreProcessing Service~\cite{Zhao_2022} & Meta's internal service for data ingestion. Replicates data pipelines across machines and wraps them behind a singular entry-point. Tailored for Meta's cluster;  adoption would require a cluster re-design to match their architecture.  \\ 
									    \midrule
\multirow{2}{*}{\textbf{\shortstack{GPU Data-Loading}}}  & DALI & Nvidia's tool for GPU-accelerated data-loading primitives, targets image processing operations (rotations, resizing, etc).  \\ 
									       & NVTabular & Nvidia's tool for GPU-accelerated data-loading primitives, focuses on tabular data. Introduces GPU resource contention between the model and data pipeline; not always practical to use.  \\ 
 \bottomrule
\label{tb:prior_art}
\end{tabularx}
\vspace{-5mm}
\end{table*}
We seek a new system --- one which can improve data-loading throughput in a general, scalable fashion without disrupting practitioner workflows or requiring large-scale cluster changes.

\vspace{2mm}

\textbf{Approach \& Contributions.} In order to reason about the data-ingestion problem from first principles, we study traces taken from our internal DLRM training cluster.
We focus on the outcomes observed by real-world DLRM practitioners, and observe shortfalls in generic data pipeline optimizers.
We study training times and processor utilization to better understand how poorly optimized data ingestion pipelines increase costs and reduce efficiency.

From our studies, we find that a \textit{lack of adaptability and feedback} is the primary missing piece in generic data pipeline optimization tools.
Out-of-memory errors, under-optimized user-defined-functions (UDFs), and poor responsiveness to dynamic machine re-sizing are the three main symptoms we observe.
Addressing the first symptom requires incorporating feedback from the system's memory usage monitor, the second requires actively adapting the optimizer's performance model of black-box UDFs, and the third requires adaptability under changing hardware conditions.

We use this key finding to motivate the design of a new data pipeline optimization tool for our cluster users at Netflix.
We build a feedback-driven, adaptive tool to optimize data ingestion pipelines that we name~\system.
\system~serves as a drop-in replacement for industry-standard optimizers such as \code{tf.data}'s \code{AUTOTUNE}, requiring no large-scale cluster redesigns or workflow disruptions.
It can be applied to any data pipeline framework, including \code{tf.data}, PyTorch Datasets, and Ray Datasets.
At the core of \system~is a \textit{reinforcement learning} (RL) agent trained on historical job traces and tuned online to understand how to distribute computational resources across the data pipeline.
RL provides the adaptability we need to mazimize performance.
The idea of a ``DL-for-systems-for-DL'' loop has recently gained traction in the systems world~\cite{virtuouscycle}; \system~provides a complete example of this loop in practice.

We apply \system~to jobs on our real-world training cluster and see 1.18-2.29X speedups in training times versus current tooling.
We observe that \system~can converge on an optimized resource distribution within only a few minutes, even on complex real-world pipelines.
Our tests show that \system~is both practical and effective in improving DLRM training efficiency.
We run scaling studies to test \system's performance further, and find that it achieves good scaling performance with respect to both workload size and machine size.

\vspace{2mm}
Our contributions can be summarized as follows:
\vspace{1mm}

\begin{enumerate}
\item We provide in-depth analyses of DLRM model training job traces taken from our real-world compute cluster, highlighting the critical and unique problem of data pipeline optimization.
\item We identify and study a new gap in the DL systems landscape, and evaluate the weaknesses of state-of-the-art tooling for data ingestion during model training.
\item We propose a novel automated data-pipeline optimizer motivated by our cluster studies, \system. To the best of our knowledge, \system~is the first system to use RL for data pipeline optimization. It is also an instance of the emerging ``DL-for-systems-for-DL'' loop.
\item We run comprehensive evaluations of \system~against state-of-the-art baselines on real-world workloads.
We find that \system~significantly outperforms the baselines by a factor of 1.18-2.29X, providing significant speedups and training cost reductions.
\end{enumerate}

The remainder of the paper is structured as follows. 
Section~\ref{sec:background} dives into the fundamentals of DL recommender training, data processing, and RL. 
Section~\ref{sec:traces} analyzes real job traces from a compute cluster to provide motivation for \system's development.
Section~\ref{sec:sys_arch} goes into the details of \system~and describes how it conceptually addresses each of the challenges we identify.
We show the results of our experimental studies in Section~\ref{sec:evaluation}, where we benchmark our system's performance on a variety of workloads.
Section~\ref{sec:prior_art} describes some existing tools for data processing and other related areas.
Finally, we provide our concluding remarks in Section~\ref{sec:conclusion}.

\section{Background}\label{sec:background}
We now provide background on the basics of DLRMs and online data pre-processing to provide context for the rest of the paper.
We then go into the basics of RL to clarify the principles that underly our proposed solution.

\subsection{Deep Learning Recommender Models}
Deep learning has becoming an increasingly popular approach to tackle recommendation problems in both industry and academia~\cite{cheng2016wide}.
These model architectures aim to bring the success that DL has seen in other domains (e.g. language modeling, object recognition) to the recommender systems space.

A typical DL model consists of a chained sequence of matrix transformations, known as layers, combined with non-linear activation functions.
The model's matrix entries, or parameters, are tuned using a historical dataset of sample-label pairs.
Each sample typically consists of multiple features, each capturing a different aspect of the historical record.
In an e-commerce dataset, for example, each record might maintain features such as item ID, the user ID, the item price, etc.
The label might be a binary indicator reflecting whether or not the user purchased the item.

The model is tuned to fit the dataset in a procedure known as stochastic gradient descent (SGD).
In SGD, batches of samples are pulled from the dataset, then fed into the model to produce predictions. 
The predictions are compared to the known ground-truth labels to produce an error value.
The derivative chain rule is used to compute a set of parameter updates to would minimize the error, which are then multiplied by a learning rate factor and applied to the model.

Recommendation data can be problematic in this context.
Tabular datasets often include categorical features (e.g. user ID, or product ID).
Such arbitrary identifiers do not carry any inherent meaning for matrix operations.
Instead, an \textit{embedding table} is used to extract meaning from these categorical identifiers.

The embedding table maps a categorical ID to a vector of continuous values.
These can be combined with any continuous sample features through an interaction procedure (e.g. concatenation).
The resultant vector can then be fed through a standard training process.
During the SGD parameter updates, the embedding vectors will be updated as though they were matrix parameters.
In theory, the embedding table is equivalent to transforming input IDs into one-hot vectors and feeding them into a standard DL model; the embedding table simply provides a more efficient representation of this same procedure.

Thus, the embedding table is the key to enabling personalized applications using DL.
Multiple major web companies now employ versions of this embedding-table DL design (e.g. Meta's DLRM~\cite{acun2021understanding} and Google's Deep \& Wide~\cite{cheng2016wide}).
It adapts the unique challenges of recommendation problems into a format amenable to DL processing.

But the embedding table introduces scaling challenges.
Consider a social media company with 1B users.
They want to be able to recommend posts to users using a DLRM model.
They select a fairly typical embedding vector size, e.g. 128.
To accommodate their 1B users, they build an embedding table with 1B entries.
Each vector entry is a 4-byte float. 
\textit{The resultant table would require more than 512GB of memory!}
DL training is typically done on a GPU to accelerate operations, but a table of this size is too large to fit into the memory of even state-of-the-art GPUs.

Various techniques have been proposed to tackle this issue, each with their own tradeoffs. 
Embedding table compression, for example, where multiple inputs map to the same vector, tends to degrade accuracy.
Another popular solution is model parallelism, where the table is sharded over multiple GPUs~\cite{wang2022merlin}.
Others have proposed hybrid compute, where the embedding table is split between system DRAM and GPU memory~\cite{sethi2022recshard,adnan2022heterogeneous}.
These scale-out solutions demand powerful and expensive hardware --- recent studies have shown that effective DLRM training requires significant infrastructure investment in both hardware and software.~\cite{acun2021understanding}.
%These are some of the largest instances offered by AWS EC2.
%Since the cost of maintaining such large machines is significant, DLRM practitioners have a vested interest in ensuring that their computational resources are used to their fullest potential.
%
%Unfortunately, this is rarely achievable with DLRMs.
%The embedding table --- the key motivator for scaling out computational demands --- is computationally inexpensive.
%Figure~\ref{fig:flops} compares the FLOPs and scales of a state-of-the-art CNN, MLP, and DLRM.

\begin{figure*}[h]
\includegraphics[width=0.8\textwidth]{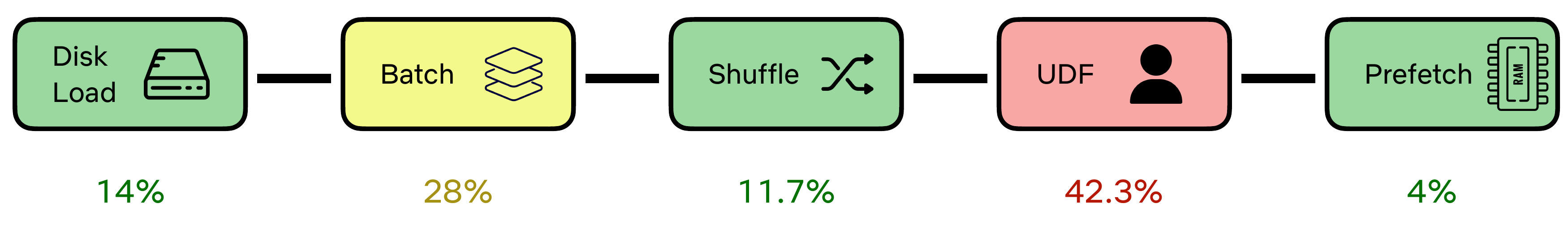}
\caption{A data processing pipeline drawn from real practice. We include the percentage of pipeline latency attributable to each stage to demonstrate the differing costs of processing.}\label{fig:pipeline}
\Description[A data pipeline illustration with associated stage latencies. The UDF stage dominates.]{Illustration demonstrating what percentage of pipeline latency is attributable to each processing stage. Disk loading takes 14\%, batching 28\%, shuffling 11.7\%, UDFs 42.3\%, and prefetching 4\%.}
\end{figure*}

%As such, both GPU-driven and CPU-driven DLRM execution tends to be very fast.
%This leaves the time-distribution of training increasingly skewed towards \textit{non-model-execution} stages such as the data processing pipeline.
%
%Consider the cost implications at Netflix, where DLRMs are trained on the GPUs offered by the p4d.24xlarge machines.
%If data processing takes up 80\% of execution times, then 80\% of GPU time is spent idly waiting for data to be received.
%The cost of the machine, then, is totally unjustified! A critical resource is being heavily underutilized.
%
%This motivates us to dig further into the data processing pipeline for DLRMs.
%Optimizing this stage of training will improve machine utilization, costs, and training times.

\subsection{DLRM Data Processing} 
The data processing challenges faced in the recommender setting differ from those seen in other domains~\cite{wang2021gradient,penedo2023refinedweb}.
A typical recommender dataset is composed of historical user interactions with the target application.
A streaming service, for example, might record user interactions (i.e. plays, ratings) with movies and shows.
\textit{Millions} of such interactions could be recorded every day.
Recommender models must be retrained regularly to account for the dataset updates.
Each model might target a different aspect of the application --- for example one model might be used for ordering rows in the UI, while another might be used for video ordering.
Individual models need different features (i.e. columns) of the base dataset and might require custom preprocessing pipelines.
Thus, it is generally impractical to push preprocessing to the offline stage; per-model customization encourages online transformation of the same base dataset.

We will now describe a typical online transformation pipeline for one such model.
Figure~\ref{fig:pipeline} illustrates, along with estimated latencies drawn from real industry pipelines.

\vspace{1mm}
Samples are loaded from the base dataset in a disk read operation.
Each sample is represented as a dictionary of key-value pairs, mapping feature names to values.
These samples are used to fill up a batch for SGD training.
This is repeated until some significant number of batches are in memory.
They are then shuffled to encourage some randomness in the SGD procedure to improve model robustness~\cite{dlbook}.

For each batch, a custom dictionary lookup operation is used to extract relevant feature columns.
In this case, we will say that product ID, user ID, user country, and total product view time are the relevant columns.
Note that this dictionary lookup could be fairly expensive on a feature-rich dataset.

The first three columns are categorical, while the fourth is continuous.
Some random noise is applied to the continuous variable to augment the data and improve model robustness.
To improve training times, several batches will be ``prefetched'' at once into GPU memory to overlap the next pipeline loading phase with model execution, trading memory for performance.
At this point, the pipeline has finished producing a training batch for model consumption.

%We now discuss the \textit{challenges} that can be observed in this pipeline.
%It should be evident from our previous descriptions that a tremendous amount of data will have to flow through the pipe.
Millions (or even billions) of recorded interactions will have to run through these stages to feed and train the model.
The throughput rate needed from this pipeline is dependent on the GPU-driven model execution speed --- in the DLRM case, this typically yields a very high rate requirement.
To improve pipeline performance, two levels of parallelism are possible.
First, pipelining.
This simply exploits the stage-by-stage processing structure.
Stages can be overlapped in a similar way to CPU instruction pipelining~\cite{crawford1990i486} to improve throughput.
Maximizing pipeline performance requires a delicate balancing act.
Each transformation stage within the pipeline must take the same amount of time to avoid idling~\cite{terapipe2021} .
%But it introduces a balancing act --- to maximize pipelining performance, each stage of processing should take roughly the same amount of time.
%Improperly aligned stage times will significantly reduce the potential gains of pipelining.
%
Second, per-stage replication. 
Replicating pipeline stages across multiple processors can improve per-stage throughput significantly.
The effect of this replication interplays with the balance of stage performances, thus impacting pipeline-parallel throughput.
Solving this complex, joint optimization problem effectively can yield significant performance benefits.
At a coarse-grained level, the entire pipeline itself could be copied across multiple machines~\cite{acun2021understanding}, but this would discard the opportunity presented by the joint optimization problem.

These challenges are complicated further by the possibility of \textit{machine resizing}.
Many clusters now use techniques such as auto-scaling, interruption \& reassignment, or even machine multi-tenancy.
In such cases, external decision-making may cause a job to actually receive \textit{new} or different resources across the course of its lifecycle.
This setting has become increasingly popular in recent years as new multi-model training tools~\cite{nagrecha2022hydra,moritz2018ray,nagrecha2021model} have emerged.
Effectively parallelizing such jobs even as the underlying resource pool is actively shifting requires a level of adaptability and flexibility not present in existing tooling.
%But there is a significant opportunity presented by fine-grained replication.
%By improving one stage or another in particular, average stage latencies can be improved such that \textit{pipeline parallel} throughput is also improved.
%
%The two levels of parallelism interplay in a complex way. 
%Properly managed stage parallelism can improve pipeline parallelism, and the two together can significantly improve training times.
%But this stage replication problem translates into a resource allocation problem.
%How many processors should be assigned to each stage for parallelism?
%And in the case of memory-driven stages such as prefetching, how much memory should be allocated?
%The problem is nontrivial, and leaving it to users to manually produce a solution is generally discouraged~\cite{murray2021tf}.
%We go into existing automated solutions (some of which are built-in to standard DL frameworks such as TensorFlow) in Section~\ref{sec:prior_art}.

\begin{figure*}[h!]
\includegraphics[width=0.95\textwidth]{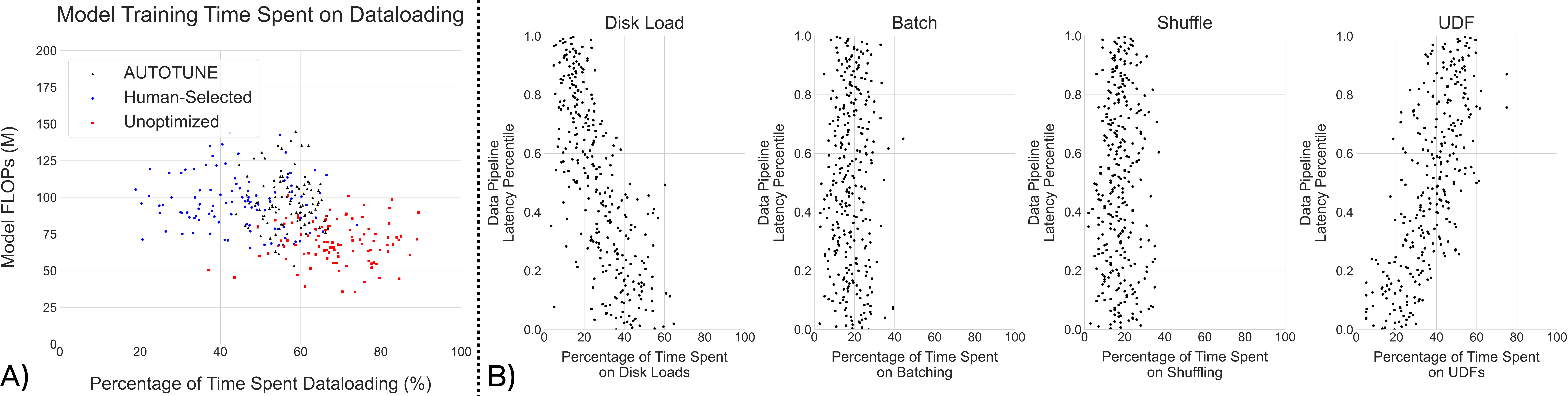}
\caption{(A) Our study of real job traces shows that compute time is dominated by data processing rather than model execution, even on the most compute-intensive models. Jobs using \code{AUTOTUNE} are marked in black, jobs using human-selected distributions are marked in blue, and unoptimized pipelines are marked in red. (B) Breakdown of individual pipeline stage latencies when using \code{AUTOTUNE}. For each stage, we provide a scatter-plot the percentage of pipeline time taken up.}\label{fig:deepdive}
\Description[Deepdive into a pipeline, illustrating that human-set approaches are better than current automated ones.]{.}\label{fig:cluster_analysis}
\end{figure*}

\subsection{Reinforcement Learning} \label{sec:rl_basics}
Next, we describe the basics of deep RL that underpin \system.

The general aim of RL is to train an ``agent'', or actor, using data collected from exploring an environment.
The agent can choose from a set of actions in the environment based on the current state.
The state is updated as a result of the action and a reward is computed to reflect the benefit produced as a result of the agent's action.
The new state and reward are used to modify the agent in a way that encourages reward-positive actions and discourages reward-negative actions.

A variety of techniques can be used to construct this feedback loop.
The Deep Q-Network (DQN) approach uses a DL model as its agent with and SGD for the feedback loop.

\textbf{DQN Technique.}
The agent model is trained to approximate an unknown function $Q$, where $Q(s, a)$ yields the reward for execution action $a$ in the environment state $state$.
Then, this DL model can compute an expected \textit{total} reward for all possible $a$'s at a given state $s$, then select the action that maximizes the expected reward.
The action space should be relatively small to make this search feasible, as excessively large action spaces are known to reduce model accuracy~\cite{thrun1993issues}.
It has become common practice to employ \textit{action space shaping}~\cite{actionspaceshaping}, reducing and combining actions to simplify the space.
In multi-discrete action spaces (e.g. a keyboard), wherein multiple simultaneous actions can be taken at once, the potential action space is exponential with a degree of the maximum number of simultaneous actions.

Selecting an action from a space requires understanding both the \textit{immediate} and \textit{long-term} reward.
To predict ``overall'' reward of an action, the Optimal Action Value Function is used~\cite{boyan1994generalization} to shape the agent's behaviors and teach it expected rewards over time.
Thus, the agent learns a model of its environment and how its actions will change its state and impact its rewards.

This design works well in settings where responsiveness and adaptability are important.
The agent can actively make decisions in response to environmental changes, a positive contrast against static one-shot optimizers.
We make use of these properties in our system design to tackle the complex and dynamic problem of data pipeline optimization.

\textbf{Online vs Offline RL.}
An RL agent can be built in either the ``offline'' or ``online'' setting.
Offline RL agents are trained in a simulation environment to understand how the various factors of their environment impact performance.
They rely on the assumption that the final, live environment will be reasonably similar to the offline simulation settings.

Online RL, by contrast, tunes the agent as it actively interacts with the target application.
This is more flexible and adaptive, but historically, long convergence times have been a significant concern.
Some recent works have proposed a hybrid of the two, initially pre-training the RL model on offline simulation data then re-tuning it online~\cite{9328612}.
The effectiveness of this hybrid is largely dependent on the specifics of the target application.

\section{Cluster Study}\label{sec:traces}
We now analyze training jobs from our real-world DLRM compute cluster to better understand the key data pipeline challenges faced by practitioners.

\subsection{Motivation}
TensorFlow, a popular DL framework, provides the \code{tf.data} API for users to build input data pipelines from the primitive operations we have discussed thus far (batch, UDF, shuffle, etc).
The new \code{torchdata} pipeline composition tool introduces similar functionality for PyTorch, though the TensorFlow data pipeline ecosystem is relatively more mature and more appropriate for our evaluations.
The \textit{de-facto} standard tool for \code{tf.data} data pipeline optimization is \code{AUTOTUNE}, which is built-in to TensorFlow~\cite{murray2021tf}.
\code{tf.data} is commonly considered to be one of the most advanced data pipeline construction tools available to practitioners, and \code{AUTOTUNE} is generally accepted to be state-of-the-art in pipeline optimization~\cite{kuchnik2022plumber}.
Due to its popularity and widespread adoption, we will take it as the standard benchmark for automated tooling in our cluster study.
Historical practice on our production cluster has surfaced three issues with \code{AUTOTUNE}.

\begin{enumerate}
	\item \textbf{Low efficiency on DLRM pipelines.} Tools like \code{AUTOTUNE} often produce suboptimal configurations in practice, bloating runtimes and costs.
	\item \textbf{High failure rates.} \code{AUTOTUNE} has shown a tendency to trigger costly out-of-memory errors, typically caused by resource-overallocations.
	\item \textbf{Poor support for rescaling.} Cluster techniques such as machine multi-tenancy or virtualization can add new resources to jobs over time. Unfortunately, \code{AUTOTUNE} does not take full advantage of the new resources without human intervention.
\end{enumerate}

We can validate these three points through quantitative analyses of DLRM job traces on our cluster.
We recorded jobs run over a period of two weeks for our study.

\begin{figure*}[h]
\includegraphics[width=\textwidth]{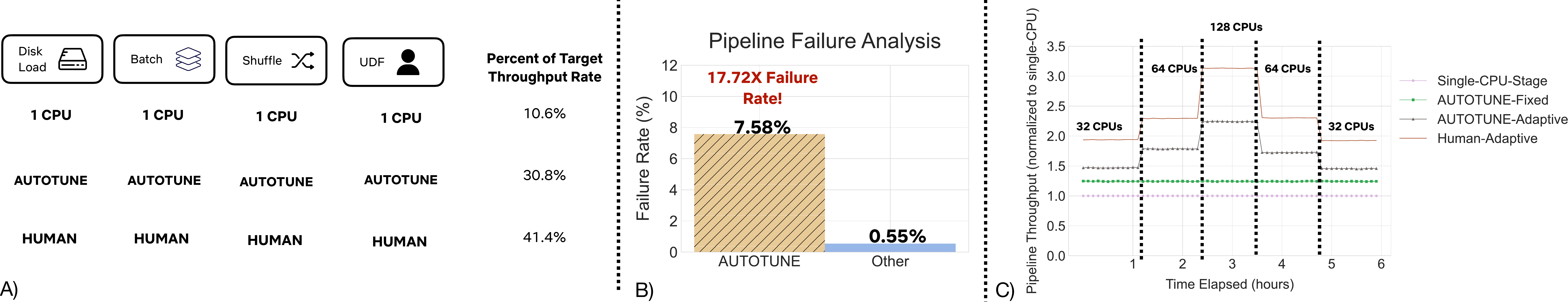}
\caption{Deepdive into a case-study pipeline. (A) The percentage of target throughput achieved with different approaches. (B) Illustration of \code{AUTOTUNE}'s versus human-set alternatives. (C) Performance of various tools when the machine is re-sized during job execution. Approaches with manual intervention use the \textit{-Adaptive} suffix, and results are normalized to the Single-CPU-per-stage baseline.}\label{fig:deepdive}
\Description[Deepdive into a pipeline, illustrating that human-set approaches are better than current automated ones.]{.}
\end{figure*}

\subsection{Cluster Trace Analyses}
We take data from a large GPU cluster reserved only for DL recommendation model training.
A broad mix of job-types are present; both exploratory ad-hoc experimentation as well as automated production pipelines.
Our results show that as much as \textit{60\%} of time is spent on data ingestion rather than model execution, even when \code{AUTOTUNE} is applied. 
Manually-optimized jobs tend to perform somewhat better, while unoptimized jobs see the worst skew towards data processing.
In all cases, we see a significant opportunity for improvement --- reducing data processing times would provide significant cost savings and efficiency improvements.
Figure~\ref{fig:cluster_analysis}(A) presents our data.

Next, we dive into the fine-grained stages of the data pipeline, to better understand the composition of the end-to-end costs.
All data pipelines in this analysis use \code{AUTOTUNE}, and follow a standard order of \code{disk load $\rightarrow$ batch $\rightarrow$ shuffle $\rightarrow$ UDF $\rightarrow$ prefetch}.
These pipelines do not include a ``cache'' stage due to memory constraints; these jobs operate with high-dimensional features \& very large datasets.
Figure~\ref{fig:cluster_analysis}(B) shows the results.

Maximizing pipeline throughput requires achieving equal latencies across each stage~\cite{terapipe2021}.
But \code{AUTOTUNE} is known to struggle with irregular stages such as UDFs or varied-size disk loads~\cite{kuchnik2022plumber}.
Our empirical study confirms this issue at a mass-scale. On average, UDF mappings and disk loading dominate runtimes, and the skew towards UDFs tends to grow as overall data pipeline latencies increase.
The proportion of time spent on shuffling or batching tends to stay mostly consistent regardless of overall pipeline latency, further pointing to UDFs as the primary stumbling-block for \code{AUTOTUNE}.
Unfortunately, UDFs are a key piece of most of DLRM data pipelines, covering basic tasks such as feature extraction, re-mapping, and transformation.
A previous study~\cite{Zhao_2022} of DLRM training describes 16 common preprocessing operations; we found that \textit{14} of these 16 required UDF implementations!
Poor UDF optimization alone is sufficient to discourage \code{AUTOTUNE} adoption among our users.

\subsection{Pipeline Deep Dive}
To gain more detailed insights, we will now analyze a singular production training pipeline from Netflix as a case study.
This job is rerun on a regular basis, multiple times per day, allowing us to collect a rich history of statistics.
Training jobs are run on machines with 128 Intel Xeon 3.0GHz CPUs, and datasets are stored on a remote network filesystem.
One of our primary aims in this section is to demonstrate how current state-of-the-art tooling fails to serve our needs.
To illustrate this, we labeled jobs according to whether they used \code{AUTOTUNE}, human-set configurations, or else no optimization at all (i.e. one CPU per stage).
We contrast these approaches in our experiments.

The model is relatively small --- under 10M parameters.
The model latency is very low, so to avoid idle times, the data pipeline must offer a high throughput rate.
The data processing pipeline requires: (1) loading data from disk, (2) shuffling it in a fixed buffer, (3) applying a UDF to extract and convert categorical features to standard mappings, then (4) batching the data before (5) prefetching it to the GPU.
If only one CPU is given to each stage, pipeline throughput is 11\% of the data-loading rate needed to keep the model served at all times (i.e. no idling), as shown in Figure~\ref{fig:deepdive}(A).
After \code{AUTOTUNE} distributes all these processors, pipeline throughput is increased by 2.81X to 31\% of the target rate.

We contrast this against \textit{manually} chosen allocations, which increases the pipeline throughput to 41\% of the target rate.
Further improvements (e.g. to 100\% of the rate, with no idle times in model execution) would require scaling beyond the machine's current resources.
But even within the machine, we found scope for a \textit{1.34X} speedup versus \code{AUTOTUNE}! Ideally, we should be able to produce this configuration automatically, without manual intervention.
%The scale of this speedup --- achieved through resource distribution alone --- is surprisingly large, but similar results have been observed in pipeline auto-tuning research for other domains~\cite{li2021terapipe}.

Another serious issue we observe in applying \code{AUTOTUNE} to this example pipeline is \textit{overallocation}. 
If we allow \code{AUTOTUNE} to take control of the \textit{prefetch} stage, it tries to improve performance by maximizing prefetches.
This bloats memory usage, often causing OOM errors.
Figure~\ref{fig:deepdive}(B) illustrates the frequency of OOM errors produced by applying \code{AUTOTUNE} to this pipeline.
Recovering from these errors requires a teardown and reset, leading to significant downtime.

An increasingly popular technique in large-scale compute clusters is machine resizing~\cite{narayanan2020heterogeneity,nagrecha2022hydra,moritz2018ray}, either from scheduler interruption \& re-assignment~\cite{xiao2018gandiva,xiao2020antman}, or job completions on a multi-tenant machine~\cite{nagrecha2022hydra}.
Rescaling the data pipeline to make use of new resources requires the optimizer to actively respond to hardware changes.
Figure~\ref{fig:deepdive}(C) charts the performance of various approaches in this setting.
We see that out-of-the-box \code{AUTOTUNE} does not respond effectively to machine re-sizing, and does not increase pipeline throughput even when new CPUs are provided to the job.
If human intervention is used to reset and re-launch \code{AUTOTUNE}, the scaling improves but is still significantly under-performs the fully-human alternative.
This reliance on human decision-making and intervention is impractical.

Now that we have observed DLRM data pipelining problems in a real training cluster, we seek a solution that can address these challenges.

\section{System Design}\label{sec:sys_arch}
Our studies in Section~\ref{sec:traces} surfaced three issues with current tooling: (1) suboptimal performance on DLRM pipelines due to UDFs, (2) inability to scale as resource caps are changed, and (3) a tendency to overallocate resources, leading to OOMs.
We propose that all three of these problems would be resolved by a \textit{feedback-driven} tool. A tool that actively evaluates its environment and collects feedback \textit{live} would be able to (1) fine-tune its understanding of UDF performance throughout training, (2) actively respond to changing resource caps, and (3) manage memory usage.

With this in mind, we turn to RL to design our tool for data pipeline optimization ---~\system.
As we discussed in Section~\ref{sec:background}, building an RL agent for data pipeline allocation requires us an environment, an agent, and an action space.
We now discuss these RL system components.

\begin{table*}[th!]
\centering
\caption{RL environment factors.}
\vspace{-4mm}
\begin{tabularx}{\textwidth}{X P{2.5cm} P{8cm} } 
& Factor & Motivation  \\
 \toprule 
\multirow{3}{*}{\textbf{\shortstack{Agent-Modified Factors}}} & Pipeline Latency & Allows agent to understand the performance of the current configuration. May change based on agent actions.  \\ 
 							       & Free CPUs & Allows agent to see how many extra CPUs it can allocate. May change based on (1) agent action or (2) machine resizing.\\ 
							       & Free Memory (bytes) & Allows agent to see how much memory it can use to increase prefetch levels. May change based on agent actions.\\ 
                                                                                \midrule
\textbf{\shortstack{Uncorrelated Factors}} & Model Latency & The actual model execution time. Updated regularly to improve estimation accuracy. Unrelated to agent actions.  \\ 
									    \midrule
\multirow{2}{*}{\textbf{\shortstack{Static Factors}}}  & DRAM-CPU Bandwidth (MB/s) & Interconnect speed can impact the value of prefetching. Found up front and unchanged throughout training.  \\ 
									       & CPU Processing Speed (GHz) & CPU processing speed can impact decision-making on resource allocation. Found up front and unchanged throughout training. \\ 
 \bottomrule
\label{tb:factors}
\end{tabularx}
\vspace{-5mm}
\end{table*}

\subsection{Environment}
The environment in our setting should reflect the data pipeline state and available hardware.
Certain aspects of the environment are static (e.g. DRAM-CPU bandwidth), others are uncorrelated to the agent's actions (e.g. model latency), while others are directly impacted by its actions (memory usage, CPU usage).
We model our environment with the aim of providing the RL agent with any and all information it might need to make an informed decision.
Our finalized list of factors is shown in in Table~\ref{tb:factors}.

\begin{figure}[h!]
\includegraphics[width=8cm]{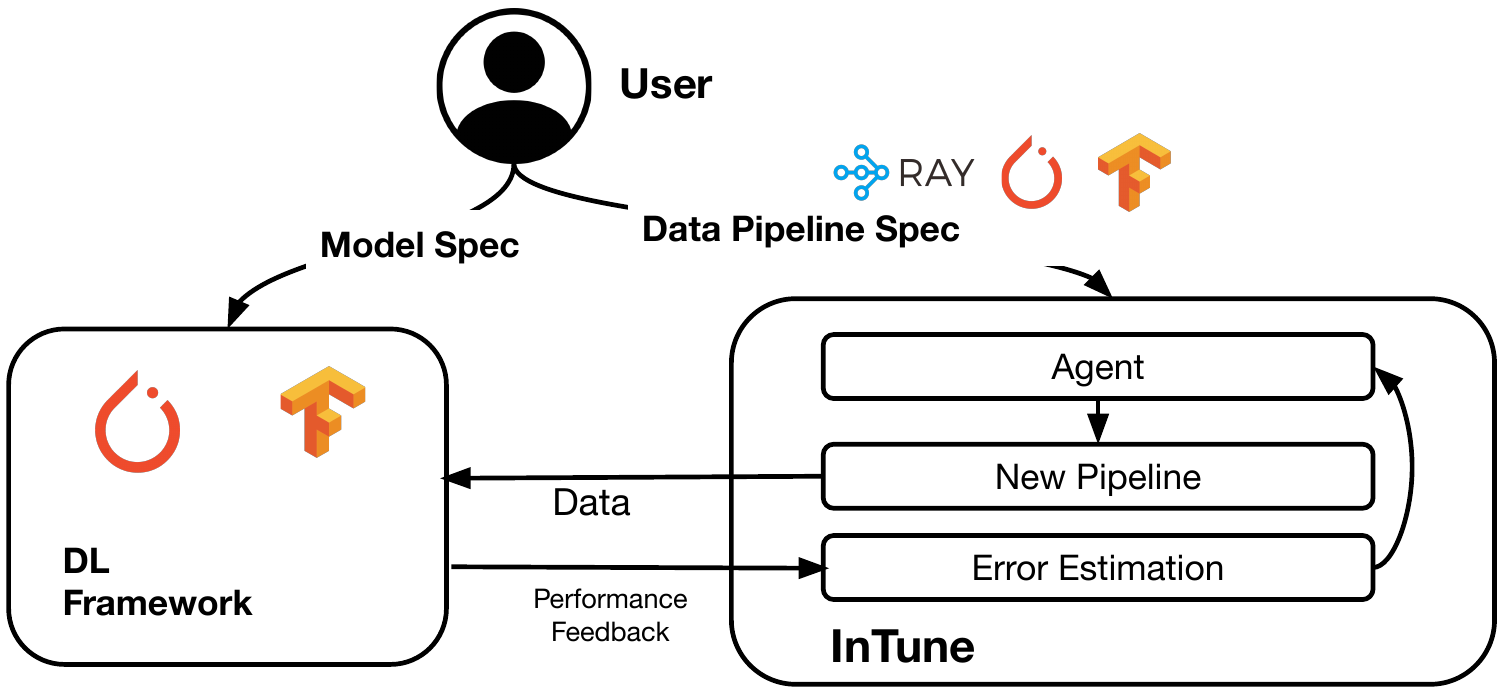}
\vspace{-4mm}
\caption{\system's RL data pipeline system architecture.}\label{fig:sys_arch}
\Description[\system's RL agent optimizes the user-specified data pipeline, which is then used to load samples for the model. The training performance is profiled and used as feedback for the agent.]{.}
%\vspace{-5mm}
\end{figure}

%
%\begin{table*}[h]\centering
%\caption{RL environment factors.}
%\label{tb:factors}
%\vspace{-4mm}
%\begin{tabular}{X @{}l p{9cm}@{}}
%\toprule
%\multicolumn{2}{c}{\textbf{Agent-Modified Factors}} \\
%\midrule
%Factor & Motivation \\
%\midrule
%Current Pipeline Latency & Allows agent to understand the performance of the current configuration. May change based on agent action. \\
%Free CPUs & Allows agent to understand how many extra CPUs it has available to allocate. May change based on (1) agent action or (2) autoscaling. \\
%Free Memory (bytes) & Allows agent to understand how much memory it has free to increase prefetches/processing levels. May change based on agent action. \\
%\midrule
%\multicolumn{2}{c}{\textbf{Variable Agent-Uncorrelated Factors}} \\
%\midrule
%Factor & Motivation \\
%\midrule
%Model Latency & The actual model execution time. Updated regularly to improve estimation accuracy. Unrelated to agent action. \\
%\midrule
%\multicolumn{2}{c}{\textbf{Static Factors}} \\
%\midrule
%Factor & Motivation \\
%\midrule
%DRAM-CPU Bandwidth (MB/s) & Interconnect speed can impact the value of prefetching. Found up front and unchanged throughout training.\\
%CPU Processing Speed (GHz) & CPU processing speed can impact decision-making on resource allocation. Found up front and unchanged throughout training.\\
%\bottomrule
%\end{tabular}
%\vspace{-4mm}
%\end{table*}

These details are sufficient for the agent to quickly grasp its problem setting.
The static factors will provide some ``immediate'' information while the other aspects will help it to learn how its actions impact data pipeline performance.
Our agent reward is directly based on data pipeline latency and memory usage. Equation~\ref{eqn:reward} shows the function.

\begin{equation}\label{eqn:reward}
R = throughput \times (1 - \frac{memory_{used}}{memory_{total}})
\end{equation}

If prefetch is not used excessively, then the memory usage portion of the equation is largely irrelevant.
But to avoid OOM outcomes like those seen with \code{AUTOTUNE}, we ensure that \system's reward approaches 0 as memory consumption nears 100\%.

\subsection{Agent Model}\label{sec:agent_model}
We aim to build a low-cost, lightweight model architecture for the RL agent.
Since \system~runs in parallel with the target DL job, we do not want to over-consume resources.
To minimize computational demands, \system's DQN agent is a simple three-layer MLP architecture using the ReLU activation function, built in PyTorch.
It can be run on either CPU or GPU resources, or even as a remote service interacting with the target job over the network.

%\begin{figure*}[h]
%\includegraphics[height=3cm]{images/mlp_arch.pdf}
%\caption{MLP model architecture.}\label{fig:model_arch}
%\end{figure*}

If the action space consists of <256 possible choices, this model only requires <200FLOPs per iteration, which should not interfere excessively with the actual model training job.
We train different versions of the agent in offline simulations to prepare them for live deployment/tuning.
Each version is built for a different common pipeline length on our clusters (e.g. one agent for 4-stage pipelines, one for 5-stage, etc).
During actual data ingestion, the model is \textit{fine-tuned} using live feedback to adapt it for the current job.
We report on convergence behaviors in Section~\ref{sec:experiments_end_to_end}.

\subsection{Action Space}
As we discussed in Section~\ref{sec:rl_basics}, it is common practice to reshape the agent action space to improve accuracy.
If we allowed our agent to directly select any distribution of resources, the size of \system's action space would be $n+r-1 \choose r-1$ where
$n$ is the number of CPUs and $r$ is the number of pipeline stages. On a typical setup (e.g. 128 CPUs over 5 stages), this would yield $1.2e7$ possible actions --- which is entirely impractical.
Based on the agent we described in Section~\ref{sec:agent_model}, this would increase iteration compute costs to more than 6.1GFLOPs!

Instead, we use action-space reshaping, and design an \textit{incremental} action space. 
At every step, \system's agent can choose to ``raise-by-one'', ``maintain'', or ``lower-by-one'' the allocation of each stage. 
Memory-bound factors use a megabyte unit while processing-bound factors use a CPU unit.
On its own, this is somewhat inefficient. 
In order for the system to allocate 128 CPUs, a minimum of $128 / n$ iterations would be required.
To improve search and convergence times, we give the agent additional options of ``raise-by-five'' and ``lower-by-five''. 
This yields an action space of $5^r$ options.
Since $r$ is typically $<=5$, this is entirely manageable. 
Increasing the action space by adding new options (e.g. ``raise-by-ten'') could be used to further modify the convergence behaviors, but we found that increment options of 1 and 5 are sufficient for \system~to rapidly converge on a performant solution.

These three components --- environment, model, and action space --- provide the basis for \system. 

\subsection{Interface \& Usage}
We aim to make \system~easy to integrate into existing user code, without disrupting workflows or requiring cluster architecture changes.
Users design their data ingestion pipelines in standard framework code (e.g. PyTorch, \code{tf.data}), then wrap their pipeline under \system, specifying any tunable performance knobs.
Performance monitoring and value adjustment is all handled automatically by \system. Listing~\ref{lst:spec} provides an example.

\begin{lstlisting}[language=Python, label={lst:spec},caption=\system~usage.,frame=single] 
pipeline = create_pipeline()
system_pipeline = intune.pipeline_wrapper(pipeline, knobs=[pipeline.stage_1, pipeline.stage_2...])
...
train(model, system_pipeline) # replace references to pipeline with system_pipeline
\end{lstlisting}

\begin{figure*}[th!]
\vspace{0mm}
\includegraphics[width=\textwidth]{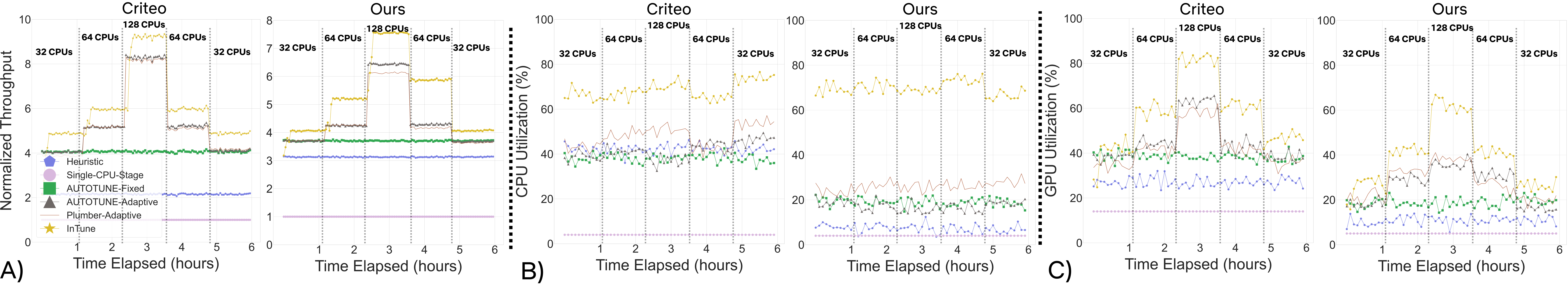}
\vspace{-5mm}
\caption{All figures use the legend in the leftmost chart. (A) Pipeline throughput over time for each approach, normalized to the Unoptimized baseline.
(B) CPU utilization over time for each approach. Only active CPUs are considered, to prevent confounding system behaviors with the separate impact of rescaling. (C) GPU Utilization over time for each approach.}
\Description[In the first figure, we demonstrate throughput on the Netflix \& Criteo pipelines. \system significantly outperforms all other baselines. In the second and third figures, we depict CPU \& GPU utilization numbers, where \system~again produces the best results.]{.}
\label{fig:e2e}
\end{figure*}

We illustrate the overall design in Figure~\ref{fig:sys_arch}.
Note the generality of \system's design; nothing about it is tied to a specific data pipeline framework. 
So long as the framework exposes optimization knobs (e.g. for CPU assignment), this approach is applicable.

\section{Evaluation}\label{sec:evaluation}

We now provide a thorough evaluation of \system.
Our aim is to answer the following questions.

\begin{enumerate}
\item Does \system~capable of achieving higher pipeline throughput than standard tools such as \code{AUTOTUNE}?
\item Is \system~less susceptible to issues such as out-of-memory errors than standard tooling?
\item Is \system~capable of responding to resource rescaling?
\item Does \system~converge on an optimized solution quickly?
\end{enumerate}

\vspace{1mm}
\textbf{Workloads.} We use two workloads, one drawn from an internal recommender model and dataset and one built using Meta's open-source DLRM code and the Criteo dataset.
The custom dataset task focuses on product recommendations while the Criteo dataset is used in a click-through-rate prediction task.

\vspace{1mm}
\textbf{Datasets and Pipelines.} The custom dataset uses dozens of sparse features, and fewer than 5 continuous features, with a batch size in the tens of thousands.
The Criteo dataset consists of 26 sparse categorical features and 13 continuous features and a batch size of 24,096.
We initialize the dataloader allocations with a simple ``even division'' of CPUs across stages.
The RL agent then modifies the distribution provided by this heuristic.
We do not consider a cache stage in the current version, since most of the relevant jobs on our cluster do not use one, but there is no reason \system~could not be extended to manage the resource allocations of cache stages as well.

\vspace{1mm}
\textbf{Models.} The model taken from our production pipeline is fairly small --- <5M parameters, most of which are contained in the embedding tables. 
We make a large model for the Criteo dataset, with 25B+ parameters, most of which are in the embedding tables. 
In both cases, the model latency is sufficiently low such that training times are dominated by data processing.

\textbf{Hardware Setup.} The production model is trained on a single 40GB A100 GPU. We initially provide the data pipeline with 32 Intel Xeon 3.0GHz CPUs, then on regular intervals double the CPU count up to a limit of 128. Then, we halve the allocation repeatedly until we reach 32 CPUs again. Approaches other than \system~rescale via manual intervention and re-launching; \system~will naturally adapt to the environmental change and so does not require this intervention step. The Criteo model is too large for a single A100 to train, so we use a standard hybrid parallelism approach~\cite{wang2022merlin} to distribute memory demands and accelerate training. We use the same CPU scaling procedure applied to the custom model (i.e. $32 \rightarrow 64 \rightarrow 128 \rightarrow 64 \rightarrow 32$). The datasets are stored on a high-bandwidth network-mounted filesystem, a common approach for large-scale recommendation datasets. 

\vspace{1mm}
\noindent \textbf{Baselines:}
We compare \system~to the following baselines.
\begin{enumerate}
	\item \textbf{Unoptimized.} In this version, only a single CPU is allocated per stage such that no parallelism is possible.
	\item \textbf{AUTOTUNE.} \code{AUTOTUNE} is a standard TensorFlow offering, commonly used to optimize \code{tf.data} pipelines.
	\item \textbf{AUTOTUNE-Adaptive.} We checkpoint and re-launch \code{AUTOTUNE} on machine rescaling intervals to allow it to adapt to the new machine resources.
	\item \textbf{Plumber-Adaptive.} Plumber~\cite{kuchnik2022plumber} is a more user-friendly alternative to \code{AUTOTUNE} that uses an MILP to distribute resources. We apply the same checkpointing approach as in \code{AUTOTUNE}-adaptive. Plumber offers additional auto-caching optimizations; we disable these since they are orthogonal to our resource-distribution work with \system~and could be integrated with \system~in the future.
	\item \textbf{Heuristic.} This version simply distributes CPUs evenly to emulate what a human user's best guess distribution might look like. \system's RL agent is initialized from this state as well.
\end{enumerate} 
These baselines cover most typical configurations, both manual and automated.

\subsection{End-to-End Performance}\label{sec:experiments_end_to_end}
\noindent \textbf{Pipeline Performance.}
We compare achieved training throughput for all approaches on both the real-world and Criteo pipelines. 
We initiate rescales at regular intervals to evaluate how each system ``responds'' to the new resource availability. 
We normalize throughput to the \textit{Unoptimized} baseline in all our analyses.
Figure~\ref{fig:e2e} (A) presents the results. 

\system~provides significantly better throughput and hardware utilization than the strongest competitors on both pipelines. 
Accounting for flexible rescaling, the average marginal gain versus standard \code{AUTOTUNE} tooling increases to 2.05X and 2.29X on the custom pipeline and Criteo pipeline respectively.
Against the alternatives which employ human intervention, the marginal improvement is still significant, ranging between 10-20\%.
Not only does our approach eliminate the headache of manual intervention, but it also achieves lower compute times and higher utilization.
In each experiment, we observe that \system~achieves a stable throughput rate within about 10 minutes.
The \textit{Plumber} baseline also requires some tens of iterations to converge, but this period is so short it does not register on the chart.
On long-running jobs, \system's ~10-minute optimization time is insignificant, but it may be problematic for short ad-hoc experiments. But in such cases, fine-tuned performance optimization is rarely important.

We also observe significantly lower \textit{failure rates} than \code{AUTOTUNE}. On average, \code{AUTOTUNE} caused an ~8\% OOM failure rate on both pipelines, whereas \system~did not cause even a single crash.
This improved robustness makes \system~an attractive option for failure-sensitive jobs.

\system also achieves higher processor utilization, illustrated in Figures~\ref{fig:e2e} (B) \& (C) likely due to reduced idling from more effective resource allocations. Some part of the CPU utilization increase can also be attributed to the overhead of maintaining a secondary RL model; unfortunately it is difficult for us to separate the two. The improved GPU utilization follows directly from the higher data throughput, since the GPU \& model are fed faster.

\noindent \textbf{Intuition on Effectiveness.} 
\system's ability to map out and tune its performance estimates over time allow it to adapt and outperform other baselines on both pipelines.
No other system can adapt as effectively to machine re-sizing out-of-the-box.
The improvements against baselines which employ human intervention can be attributed to one of the other primary weaknesses we observed in Section~\ref{sec:traces}, UDF performance modeling.
As we will show in Section~\ref{sec:complex_scale}, \system~proves to be significantly better than the strongest baseline --- \code{AUTOTUNE} --- in optimizing UDF pipeline stages.

\subsection{Drilldown Studies}\label{sec:ablation}

We now dive into \system's scaling behaviors on the real-world data pipeline.
We aim to answer the following questions.

\begin{enumerate}
	\item How does \system's performance change as the pipeline's complexity is changed?
	\item How does \system's performance change as CPU counts are changed?
	\item How does \system's performance change as batch size changes?
\end{enumerate}

\begin{figure*}[th!]
\vspace{0mm}
\includegraphics[height=3cm]{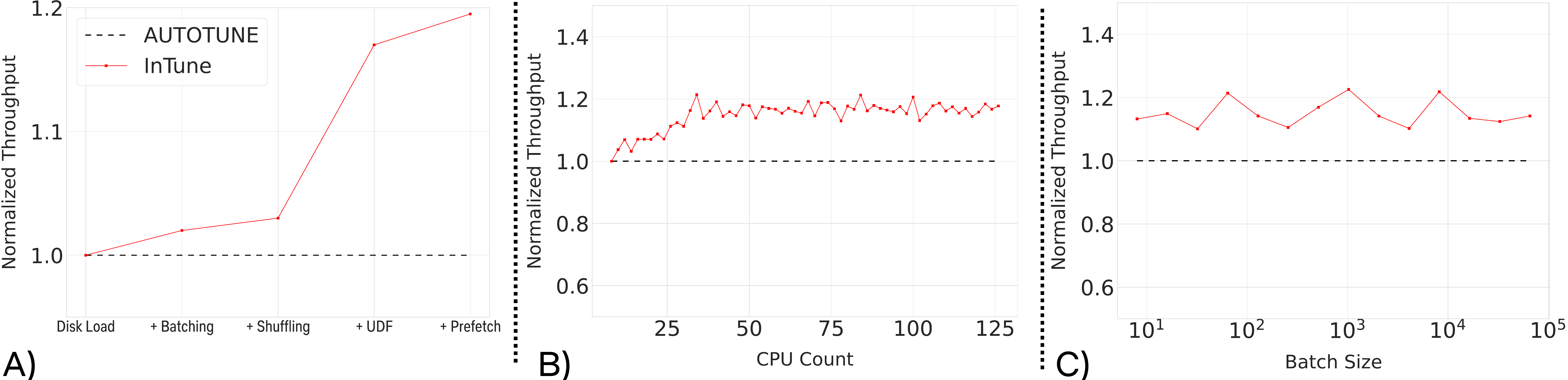}
\caption{Performance scaling with respect to (A) pipeline complexity, (B) CPU count, and (C) batch size.}
\label{fig:scaling}
\Description[We determine how \system's performance scales with pipeline complexity, CPU count, and batch size. In general, we find that \system's marginal improvements over baselines like \code{AUTOTUNE} scale up with pipeline complexity and are relatively consistent versus CPU count and batch size.]{.}
\vspace{-3mm}
\end{figure*}

\subsubsection{Pipeline Complexity Scaling}\label{sec:complex_scale}
We report on performance normalized to the \code{AUTOTUNE} baseline on the same pipeline.
All settings use the same machine, with 128 Intel Xeon 3.0GHz CPUs, and a constant model latency of 0s (to encourage maximal pipeline optimization).
Pipeline ``complexity'' is adjusted by increasing/decreasing pipeline length (e.g. \code{+ batching}, \code{+ shuffling}).
Figure~\ref{fig:scaling}(A) presents the results.
We see that our system's marginal improvement over the \code{AUTOTUNE} baseline grows as pipeline complexity increases, with a spike when UDFs are introduced.
This corroborates earlier studies which found that \code{AUTOTUNE} underperforms on more complex, UDF-driven pipelines~\cite{kuchnik2022plumber}.

\subsubsection{Machine Size Scaling}
We now study the scaling efficacy of \system. 
We increase CPU count in increments of 2, ranging from 8 to 128.
\code{AUTOTUNE} is re-launched at each machine size to rebase the relative performance.
Figure~\ref{fig:scaling}(B) presents the results.
\system's relative improvements over \code{AUTOTUNE} tend to grow as the valid configuration space increases, but then flattens out to a constant outperformance of roughly 20\%.
This flattening should be expected; \code{AUTOTUNE} is a strong baseline and even a 20\% performance margin is significant~\cite{kuchnik2022plumber}.

\subsubsection{Batch Size Scaling}
In our final scaling study, we evaluate our system's performance with respect to batch size. 
Like the pipeline complexity study, we evaluate our system's ability to respond to varied workload intensity.
Larger batch sizes increase demands on specific stages (i.e. the batch stage, prefetch stage, and possibly the UDF stage).
Since end-system users might wish to deploy the system on any range of batch sizes, we implement this study to give a more thorough understanding of our system's offerings to users.
Figure~\ref{fig:scaling}(C) presents the results.
We see that our system manages to maintain (and even improve) average sample throughput even as batch size increases.

\section{Related Work}\label{sec:prior_art}
We now discuss related work \& prior art.
We can divide these works into three major categories: data pipeline tools, DL resource allocation tools, and works on RL for systems.

%\vspace{1mm}
\subsection{Data Pipeline Optimizers}
\textbf{\code{AUTOTUNE}} is generally considered to be the gold-standard data pipeline optimizer for \code{tf.data} pipelines~\cite{murray2021tf}. 
Built-in to TensorFlow, the tool offers users with a seamless optimization experience that can be added to existing pipelines in just a single line of code.
This design philosophy of abstraction has its detractors; recent works~\cite{kuchnik2022plumber} have criticized its black-box approach to optimization.
\code{AUTOTUNE} is designed to support any \code{tf.data} pipeline, but its generality leaves it vulnerable to task-specific issues, such as those we outline in Section~\ref{sec:intro} and explore in Section~\ref{sec:traces}.
\system~is more narrowly focused than \code{AUTOTUNE} in target workloads, but is also more general in its support for non-\code{tf.data} pipelines.

\textbf{Plumber} was introduced as a more user-friendly alternative to \code{AUTOTUNE} (but still restricted to \code{tf.data}). It uses a linear programming solver to determine a resource allocation.
However, in practice it often \textit{underperforms} \code{AUTOTUNE}~\cite{kuchnik2022plumber} in its allocations. It does offer the ability to automatically
inject caching into the pipeline to improve performance. A future version of our system could borrow this optimization from Plumber and add caching as an action for the agent.

\textbf{DALI \& NVTabular} offer GPU-accelerated data-loader primitives for image and tabular data modalities respectively
In practice, we find that NVTabular is more suited to offline feature engineering, since using the GPU for online data ingestion can lead to contention over cycles between the pipeline and the model.
As a result, practitioners on our cluster have generally found it impractical to adopt NVTabular for our target setting.

\textbf{CoorDL} proposed a set of carefully-designed techniques to eliminate data stalls, including sophisticated caching procedures~\cite{mohan2020analyzing}. 
We address the related, but orthogonal, problem of data pipeline parallelization and throughput optimization.
We leave it to future work to combine their results with our own. Another recent work~\cite{Isenko_2022} focused on analyzing various training pipelines and identifying typical bottlenecks.
They characterize general pipeline design spaces; by contrast, our work focuses specifically on DLRM challenges.

Meta's \textbf{Data PreProcessing Service}~\cite{Zhao_2022} tackles a similar problem to ours --- online data ingestion pipelines for large-scale DLRM training.
Their work primarily focuses on understanding performance issues in Meta's cluster, and describing their ``disaggregated data service'' approach.
The idea of the service is to place replicas of the data ingestion pipeline on additional, separate CPU servers, which feed the trainer machine the data samples over the network.
The opportunity to scale beyond trainer machine CPUs is attractive, but adopting this approach requires significant cluster redesigns, altering user workflows, and creating new machine scheduling schemes.
In addition, replicating the entire pipeline over multiple machines can be wasteful if the bottleneck is only at a single processing stage.
By contrast, our work focuses on maximizing the effectiveness of CPUs already present on the trainer machine, so that no alterations to the user workflow or cluster setup are necessary.
In theory, the two approaches are orthogonal --- our CPU-distribution scheme could be applied to each of the machine replicas used in Meta's disaggregated data-loading design.

Other query optimization tools~\cite{nakandala2020incremental,kumar2021cerebro,nakandala2022nautilus,huang2019gpipe,wang2021sensai} exist, but target settings other than data pipeline optimization.

%\vspace{1mm}
\subsection{DL Resource Allocation}
Some works (e.g. Saturn~\cite{nagrechasaturn}, Pollux~\cite{qiao2021pollux}, Optimus~\cite{peng2018optimus} and DL2~\cite{peng2021dl2}) have tackled GPU apportioning in the scheduling setting. Still other works~\cite{li2018massively} consider general resource apportionment for hyperparameter tuning. Like \system, these tools reduce the manual configuration burden in the DL training process.

%\vspace{1mm}
\subsection{Deep Reinforcement Learning for Systems}
Deep RL has become increasingly popular in recent years for various systems applications~\cite{ameer2019view}.
Several works have tackled resource allocation using RL~\cite{9328612,8198798,Tesauro2005OnlinePM}.
They aim to use the flexibility of RL to tackle the complexity and dynamic nature of intractable, online problems.
Similarly, our work exploits the flexibility of RL to meet the needs of recommender data ingestion pipelines that are unaddressed by existing systems.
Others have applied RL for SQL optimization~\cite{zhong2017seq,Marcus_2019,Marcus_2018,ortiz2018learning,krishnan2018learning}.
The use of a learned algorithm helps relax the need for exact information that may be impractical to obtain in large RDBMSs.
Our work also uses RL to relax the need for exact profiling of blackbox UDFs.

\section{Conclusion}\label{sec:conclusion}
DLRM training costs are often dominated by online data ingestion rather than model execution.
The primary throughput bottleneck in this setting is \textit{CPU-driven} data processing rather than \textit{GPU-accelerated} model operations.
Thus, optimizing the data ingestion phase is critical to ensuring cost- \& time- effective model development.
Unfortunately, existing tooling for DL data ingestion pipelines does not support DLRM setting effectively.
We draw on lessons learned from analyses of real DLRM workloads from our training cluster at Netflix to motivate the design of a novel RL-based system we name \system.
\system~dynamically allocates CPUs \& memory across stages of the online data ingestion pipeline, significantly improving efficiency over industry-standard baselines without requiring changes to the cluster architecture or user workflows. 
Benchmarks on real \& synthetic training pipelines show that our system outperforms the strongest out-of-the-box tools by >2X, and human-managed baselines by up to 20\%.
Overall, \system~offers significant performance and cost benefits for recommender training pipelines and should serve to encourage further training optimization works customized for the unique needs of DLRM training.
Future extensions could extend \system~to other decisions in the DLRM data pipeline; e.g. intermediate caching, or how to scale the pipeline across multiple machines.

\bibliographystyle{ACM-Reference-Format}
\bibliography{recsys}


\begin{thebibliography}{55}

%%% the Web address.

\ifx \showCODEN    \undefined \def \showCODEN     #1{\unskip}     \fi
\ifx \showDOI      \undefined \def \showDOI       #1{#1}\fi
\ifx \showISBNx    \undefined \def \showISBNx     #1{\unskip}     \fi
\ifx \showISBNxiii \undefined \def \showISBNxiii  #1{\unskip}     \fi
\ifx \showISSN     \undefined \def \showISSN      #1{\unskip}     \fi
\ifx \showLCCN     \undefined \def \showLCCN      #1{\unskip}     \fi
\ifx \shownote     \undefined \def \shownote      #1{#1}          \fi
\ifx \showarticletitle \undefined \def \showarticletitle #1{#1}   \fi
\ifx \showURL      \undefined \def \showURL       {\relax}        \fi
% invisible to TeX
\providecommand\bibfield[2]{#2}
\providecommand\bibinfo[2]{#2}
\providecommand\natexlab[1]{#1}
\providecommand\showeprint[2][]{arXiv:#2}

\bibitem[Acun et~al\mbox{.}(2021)]%
        {acun2021understanding}
\bibfield{author}{\bibinfo{person}{Bilge Acun}, \bibinfo{person}{Matthew
  Murphy}, \bibinfo{person}{Xiaodong Wang}, \bibinfo{person}{Jade Nie},
  \bibinfo{person}{Carole-Jean Wu}, {and} \bibinfo{person}{Kim Hazelwood}.}
  \bibinfo{year}{2021}\natexlab{}.
\newblock \showarticletitle{Understanding training efficiency of deep learning
  recommendation models at scale}. In \bibinfo{booktitle}{\emph{2021 IEEE
  International Symposium on High-Performance Computer Architecture (HPCA)}}.
  IEEE, \bibinfo{pages}{802--814}.
\newblock


\bibitem[Adnan et~al\mbox{.}(2022)]%
        {adnan2022heterogeneous}
\bibfield{author}{\bibinfo{person}{Muhammad Adnan},
  \bibinfo{person}{Yassaman~Ebrahimzadeh Maboud}, \bibinfo{person}{Divya
  Mahajan}, {and} \bibinfo{person}{Prashant~J Nair}.}
  \bibinfo{year}{2022}\natexlab{}.
\newblock \showarticletitle{Heterogeneous Acceleration Pipeline for
  Recommendation System Training}.
\newblock \bibinfo{journal}{\emph{arXiv preprint arXiv:2204.05436}}
  (\bibinfo{year}{2022}).
\newblock


\bibitem[Boyan and Moore(1994)]%
        {boyan1994generalization}
\bibfield{author}{\bibinfo{person}{Justin Boyan} {and} \bibinfo{person}{Andrew
  Moore}.} \bibinfo{year}{1994}\natexlab{}.
\newblock \showarticletitle{Generalization in reinforcement learning: Safely
  approximating the value function}.
\newblock \bibinfo{journal}{\emph{Advances in neural information processing
  systems}}  \bibinfo{volume}{7} (\bibinfo{year}{1994}).
\newblock


\bibitem[Cheng et~al\mbox{.}(2016)]%
        {cheng2016wide}
\bibfield{author}{\bibinfo{person}{Heng-Tze Cheng}, \bibinfo{person}{Levent
  Koc}, \bibinfo{person}{Jeremiah Harmsen}, \bibinfo{person}{Tal Shaked},
  \bibinfo{person}{Tushar Chandra}, \bibinfo{person}{Hrishi Aradhye},
  \bibinfo{person}{Glen Anderson}, \bibinfo{person}{Greg Corrado},
  \bibinfo{person}{Wei Chai}, \bibinfo{person}{Mustafa Ispir}, {et~al\mbox{.}}}
  \bibinfo{year}{2016}\natexlab{}.
\newblock \showarticletitle{Wide \& deep learning for recommender systems}. In
  \bibinfo{booktitle}{\emph{Proceedings of the 1st workshop on deep learning
  for recommender systems}}. \bibinfo{pages}{7--10}.
\newblock


\bibitem[Crawford(1990)]%
        {crawford1990i486}
\bibfield{author}{\bibinfo{person}{John~H Crawford}.}
  \bibinfo{year}{1990}\natexlab{}.
\newblock \showarticletitle{The i486 CPU: Executing instructions in one clock
  cycle}.
\newblock \bibinfo{journal}{\emph{IEEE Micro}} \bibinfo{volume}{10},
  \bibinfo{number}{1} (\bibinfo{year}{1990}), \bibinfo{pages}{27--36}.
\newblock


\bibitem[Goodfellow et~al\mbox{.}(2016)]%
        {dlbook}
\bibfield{author}{\bibinfo{person}{Ian Goodfellow}, \bibinfo{person}{Yoshua
  Bengio}, {and} \bibinfo{person}{Aaron Courville}.}
  \bibinfo{year}{2016}\natexlab{}.
\newblock \bibinfo{booktitle}{\emph{Deep Learning}}.
\newblock \bibinfo{publisher}{MIT Press}.
\newblock
\newblock
\shownote{\url{http://www.deeplearningbook.org}}.


\bibitem[Gupta et~al\mbox{.}(2020)]%
        {gupta2020architectural}
\bibfield{author}{\bibinfo{person}{Udit Gupta}, \bibinfo{person}{Carole-Jean
  Wu}, \bibinfo{person}{Xiaodong Wang}, \bibinfo{person}{Maxim Naumov},
  \bibinfo{person}{Brandon Reagen}, \bibinfo{person}{David Brooks},
  \bibinfo{person}{Bradford Cottel}, \bibinfo{person}{Kim Hazelwood},
  \bibinfo{person}{Mark Hempstead}, \bibinfo{person}{Bill Jia},
  {et~al\mbox{.}}} \bibinfo{year}{2020}\natexlab{}.
\newblock \showarticletitle{The architectural implications of facebook's
  dnn-based personalized recommendation}. In \bibinfo{booktitle}{\emph{2020
  IEEE International Symposium on High Performance Computer Architecture
  (HPCA)}}. IEEE, \bibinfo{pages}{488--501}.
\newblock


\bibitem[Haj-Ali et~al\mbox{.}(2019)]%
        {ameer2019view}
\bibfield{author}{\bibinfo{person}{Ameer Haj-Ali}, \bibinfo{person}{Nesreen~K.
  Ahmed}, \bibinfo{person}{Ted Willke}, \bibinfo{person}{Joseph Gonzalez},
  \bibinfo{person}{Krste Asanovic}, {and} \bibinfo{person}{Ion Stoica}.}
  \bibinfo{year}{2019}\natexlab{}.
\newblock \bibinfo{title}{A View on Deep Reinforcement Learning in System
  Optimization}.
\newblock
\newblock
\urldef\tempurl%
\url{https://doi.org/10.48550/ARXIV.1908.01275}
\showDOI{\tempurl}


\bibitem[Harlap et~al\mbox{.}(2018)]%
        {pipedream2018}
\bibfield{author}{\bibinfo{person}{Aaron Harlap}, \bibinfo{person}{Deepak
  Narayanan}, \bibinfo{person}{Amar Phanishayee}, \bibinfo{person}{Vivek
  Seshadri}, \bibinfo{person}{Nikhil Devanur}, \bibinfo{person}{Greg Ganger},
  {and} \bibinfo{person}{Phil Gibbons}.} \bibinfo{year}{2018}\natexlab{}.
\newblock \bibinfo{title}{PipeDream: Fast and Efficient Pipeline Parallel DNN
  Training}.
\newblock
\newblock
\urldef\tempurl%
\url{https://doi.org/10.48550/ARXIV.1806.03377}
\showDOI{\tempurl}


\bibitem[He et~al\mbox{.}(2017)]%
        {8198798}
\bibfield{author}{\bibinfo{person}{Ying He}, \bibinfo{person}{F.~Richard Yu},
  \bibinfo{person}{Nan Zhao}, \bibinfo{person}{Victor C.~M. Leung}, {and}
  \bibinfo{person}{Hongxi Yin}.} \bibinfo{year}{2017}\natexlab{}.
\newblock \showarticletitle{Software-Defined Networks with Mobile Edge
  Computing and Caching for Smart Cities: A Big Data Deep Reinforcement
  Learning Approach}.
\newblock \bibinfo{journal}{\emph{IEEE Communications Magazine}}
  \bibinfo{volume}{55}, \bibinfo{number}{12} (\bibinfo{year}{2017}),
  \bibinfo{pages}{31--37}.
\newblock


\bibitem[Huang et~al\mbox{.}(2019)]%
        {huang2019gpipe}
\bibfield{author}{\bibinfo{person}{Yanping Huang}, \bibinfo{person}{Youlong
  Cheng}, \bibinfo{person}{Ankur Bapna}, \bibinfo{person}{Orhan Firat},
  \bibinfo{person}{Dehao Chen}, \bibinfo{person}{Mia Chen},
  \bibinfo{person}{HyoukJoong Lee}, \bibinfo{person}{Jiquan Ngiam},
  \bibinfo{person}{Quoc~V Le}, \bibinfo{person}{Yonghui Wu}, {et~al\mbox{.}}}
  \bibinfo{year}{2019}\natexlab{}.
\newblock \showarticletitle{Gpipe: Efficient training of giant neural networks
  using pipeline parallelism}.
\newblock \bibinfo{journal}{\emph{Advances in neural information processing
  systems}}  \bibinfo{volume}{32} (\bibinfo{year}{2019}).
\newblock


\bibitem[Huang et~al\mbox{.}(2018)]%
        {gpipe2018}
\bibfield{author}{\bibinfo{person}{Yanping Huang}, \bibinfo{person}{Youlong
  Cheng}, \bibinfo{person}{Ankur Bapna}, \bibinfo{person}{Orhan Firat},
  \bibinfo{person}{Mia~Xu Chen}, \bibinfo{person}{Dehao Chen},
  \bibinfo{person}{HyoukJoong Lee}, \bibinfo{person}{Jiquan Ngiam},
  \bibinfo{person}{Quoc~V. Le}, \bibinfo{person}{Yonghui Wu}, {and}
  \bibinfo{person}{Zhifeng Chen}.} \bibinfo{year}{2018}\natexlab{}.
\newblock \bibinfo{title}{GPipe: Efficient Training of Giant Neural Networks
  using Pipeline Parallelism}.
\newblock
\newblock
\urldef\tempurl%
\url{https://doi.org/10.48550/ARXIV.1811.06965}
\showDOI{\tempurl}


\bibitem[Isenko et~al\mbox{.}(2022)]%
        {Isenko_2022}
\bibfield{author}{\bibinfo{person}{Alexander Isenko}, \bibinfo{person}{Ruben
  Mayer}, \bibinfo{person}{Jeffrey Jedele}, {and} \bibinfo{person}{Hans-Arno
  Jacobsen}.} \bibinfo{year}{2022}\natexlab{}.
\newblock \showarticletitle{Where Is My Training Bottleneck? Hidden Trade-Offs
  in Deep Learning Preprocessing Pipelines}. In
  \bibinfo{booktitle}{\emph{Proceedings of the 2022 International Conference on
  Management of Data}}. \bibinfo{publisher}{{ACM}}.
\newblock
\urldef\tempurl%
\url{https://doi.org/10.1145/3514221.3517848}
\showDOI{\tempurl}


\bibitem[Jia et~al\mbox{.}(2018)]%
        {flexflow2018}
\bibfield{author}{\bibinfo{person}{Zhihao Jia}, \bibinfo{person}{Matei
  Zaharia}, {and} \bibinfo{person}{Alex Aiken}.}
  \bibinfo{year}{2018}\natexlab{}.
\newblock \bibinfo{title}{Beyond Data and Model Parallelism for Deep Neural
  Networks}.
\newblock
\newblock
\urldef\tempurl%
\url{https://doi.org/10.48550/ARXIV.1807.05358}
\showDOI{\tempurl}


\bibitem[Kanervisto et~al\mbox{.}(2020)]%
        {actionspaceshaping}
\bibfield{author}{\bibinfo{person}{Anssi Kanervisto},
  \bibinfo{person}{Christian Scheller}, {and} \bibinfo{person}{Ville
  Hautamäki}.} \bibinfo{year}{2020}\natexlab{}.
\newblock \bibinfo{title}{Action Space Shaping in Deep Reinforcement Learning}.
\newblock
\newblock
\urldef\tempurl%
\url{https://doi.org/10.48550/ARXIV.2004.00980}
\showDOI{\tempurl}


\bibitem[Krishnan et~al\mbox{.}(2018)]%
        {krishnan2018learning}
\bibfield{author}{\bibinfo{person}{Sanjay Krishnan}, \bibinfo{person}{Zongheng
  Yang}, \bibinfo{person}{Ken Goldberg}, \bibinfo{person}{Joseph Hellerstein},
  {and} \bibinfo{person}{Ion Stoica}.} \bibinfo{year}{2018}\natexlab{}.
\newblock \bibinfo{title}{Learning to Optimize Join Queries With Deep
  Reinforcement Learning}.
\newblock
\newblock
\urldef\tempurl%
\url{https://doi.org/10.48550/ARXIV.1808.03196}
\showDOI{\tempurl}


\bibitem[Kuchnik et~al\mbox{.}(2022)]%
        {kuchnik2022plumber}
\bibfield{author}{\bibinfo{person}{Michael Kuchnik}, \bibinfo{person}{Ana
  Klimovic}, \bibinfo{person}{Jiri Simsa}, \bibinfo{person}{Virginia Smith},
  {and} \bibinfo{person}{George Amvrosiadis}.} \bibinfo{year}{2022}\natexlab{}.
\newblock \showarticletitle{Plumber: Diagnosing and removing performance
  bottlenecks in machine learning data pipelines}.
\newblock \bibinfo{journal}{\emph{Proceedings of Machine Learning and Systems}}
   \bibinfo{volume}{4} (\bibinfo{year}{2022}), \bibinfo{pages}{33--51}.
\newblock


\bibitem[Kumar et~al\mbox{.}(2021)]%
        {kumar2021cerebro}
\bibfield{author}{\bibinfo{person}{Arun Kumar}, \bibinfo{person}{Supun
  Nakandala}, \bibinfo{person}{Yuhao Zhang}, \bibinfo{person}{Side Li},
  \bibinfo{person}{Advitya Gemawat}, {and} \bibinfo{person}{Kabir Nagrecha}.}
  \bibinfo{year}{2021}\natexlab{}.
\newblock \showarticletitle{Cerebro: A layered data platform for scalable deep
  learning}. In \bibinfo{booktitle}{\emph{11th Annual Conference on Innovative
  Data Systems Research (CIDR ‘21)}}.
\newblock


\bibitem[Lepikhin et~al\mbox{.}(2020)]%
        {gshard2020}
\bibfield{author}{\bibinfo{person}{Dmitry Lepikhin},
  \bibinfo{person}{HyoukJoong Lee}, \bibinfo{person}{Yuanzhong Xu},
  \bibinfo{person}{Dehao Chen}, \bibinfo{person}{Orhan Firat},
  \bibinfo{person}{Yanping Huang}, \bibinfo{person}{Maxim Krikun},
  \bibinfo{person}{Noam Shazeer}, {and} \bibinfo{person}{Zhifeng Chen}.}
  \bibinfo{year}{2020}\natexlab{}.
\newblock \bibinfo{title}{GShard: Scaling Giant Models with Conditional
  Computation and Automatic Sharding}.
\newblock
\newblock
\urldef\tempurl%
\url{https://doi.org/10.48550/ARXIV.2006.16668}
\showDOI{\tempurl}


\bibitem[Li et~al\mbox{.}(2018)]%
        {li2018massively}
\bibfield{author}{\bibinfo{person}{Liam Li}, \bibinfo{person}{Kevin Jamieson},
  \bibinfo{person}{Afshin Rostamizadeh}, \bibinfo{person}{Ekaterina Gonina},
  \bibinfo{person}{Moritz Hardt}, \bibinfo{person}{Benjamin Recht}, {and}
  \bibinfo{person}{Ameet Talwalkar}.} \bibinfo{year}{2018}\natexlab{}.
\newblock \showarticletitle{Massively parallel hyperparameter tuning}.
\newblock \bibinfo{journal}{\emph{arXiv preprint arXiv:1810.05934}}
  \bibinfo{volume}{5} (\bibinfo{year}{2018}).
\newblock


\bibitem[Li et~al\mbox{.}(2021)]%
        {terapipe2021}
\bibfield{author}{\bibinfo{person}{Zhuohan Li}, \bibinfo{person}{Siyuan
  Zhuang}, \bibinfo{person}{Shiyuan Guo}, \bibinfo{person}{Danyang Zhuo},
  \bibinfo{person}{Hao Zhang}, \bibinfo{person}{Dawn Song}, {and}
  \bibinfo{person}{Ion Stoica}.} \bibinfo{year}{2021}\natexlab{}.
\newblock \bibinfo{title}{TeraPipe: Token-Level Pipeline Parallelism for
  Training Large-Scale Language Models}.
\newblock
\newblock
\urldef\tempurl%
\url{https://doi.org/10.48550/ARXIV.2102.07988}
\showDOI{\tempurl}


\bibitem[Marcus et~al\mbox{.}(2019)]%
        {Marcus_2019}
\bibfield{author}{\bibinfo{person}{Ryan Marcus}, \bibinfo{person}{Parimarjan
  Negi}, \bibinfo{person}{Hongzi Mao}, \bibinfo{person}{Chi Zhang},
  \bibinfo{person}{Mohammad Alizadeh}, \bibinfo{person}{Tim Kraska},
  \bibinfo{person}{Olga Papaemmanouil}, {and} \bibinfo{person}{Nesime Tatbul}.}
  \bibinfo{year}{2019}\natexlab{}.
\newblock \showarticletitle{Neo}.
\newblock \bibinfo{journal}{\emph{Proceedings of the {VLDB} Endowment}}
  \bibinfo{volume}{12}, \bibinfo{number}{11} (\bibinfo{date}{jul}
  \bibinfo{year}{2019}), \bibinfo{pages}{1705--1718}.
\newblock
\urldef\tempurl%
\url{https://doi.org/10.14778/3342263.3342644}
\showDOI{\tempurl}


\bibitem[Marcus and Papaemmanouil(2018)]%
        {Marcus_2018}
\bibfield{author}{\bibinfo{person}{Ryan Marcus} {and} \bibinfo{person}{Olga
  Papaemmanouil}.} \bibinfo{year}{2018}\natexlab{}.
\newblock \showarticletitle{Deep Reinforcement Learning for Join Order
  Enumeration}. In \bibinfo{booktitle}{\emph{Proceedings of the First
  International Workshop on Exploiting Artificial Intelligence Techniques for
  Data Management}}. \bibinfo{publisher}{{ACM}}.
\newblock
\urldef\tempurl%
\url{https://doi.org/10.1145/3211954.3211957}
\showDOI{\tempurl}


\bibitem[Mohan et~al\mbox{.}(2020)]%
        {mohan2020analyzing}
\bibfield{author}{\bibinfo{person}{Jayashree Mohan}, \bibinfo{person}{Amar
  Phanishayee}, \bibinfo{person}{Ashish Raniwala}, {and} \bibinfo{person}{Vijay
  Chidambaram}.} \bibinfo{year}{2020}\natexlab{}.
\newblock \bibinfo{title}{Analyzing and Mitigating Data Stalls in DNN
  Training}.
\newblock
\newblock
\urldef\tempurl%
\url{https://doi.org/10.48550/ARXIV.2007.06775}
\showDOI{\tempurl}


\bibitem[Moritz et~al\mbox{.}(2018)]%
        {moritz2018ray}
\bibfield{author}{\bibinfo{person}{Philipp Moritz}, \bibinfo{person}{Robert
  Nishihara}, \bibinfo{person}{Stephanie Wang}, \bibinfo{person}{Alexey
  Tumanov}, \bibinfo{person}{Richard Liaw}, \bibinfo{person}{Eric Liang},
  \bibinfo{person}{Melih Elibol}, \bibinfo{person}{Zongheng Yang},
  \bibinfo{person}{William Paul}, \bibinfo{person}{Michael~I. Jordan}, {and}
  \bibinfo{person}{Ion Stoica}.} \bibinfo{year}{2018}\natexlab{}.
\newblock \bibinfo{title}{Ray: A Distributed Framework for Emerging AI
  Applications}.
\newblock
\newblock
\showeprint[arxiv]{1712.05889}~[cs.DC]


\bibitem[Mudigere et~al\mbox{.}(2021)]%
        {mudigere2021software}
\bibfield{author}{\bibinfo{person}{Dheevatsa Mudigere} {et~al\mbox{.}}}
  \bibinfo{year}{2021}\natexlab{}.
\newblock \bibinfo{title}{Software-Hardware Co-design for Fast and Scalable
  Training of Deep Learning Recommendation Models}.
\newblock
\newblock
\urldef\tempurl%
\url{https://doi.org/10.48550/ARXIV.2104.05158}
\showDOI{\tempurl}


\bibitem[Murray et~al\mbox{.}(2021)]%
        {murray2021tf}
\bibfield{author}{\bibinfo{person}{Derek~G Murray}, \bibinfo{person}{Jiri
  Simsa}, \bibinfo{person}{Ana Klimovic}, {and} \bibinfo{person}{Ihor Indyk}.}
  \bibinfo{year}{2021}\natexlab{}.
\newblock \showarticletitle{tf. data: A machine learning data processing
  framework}.
\newblock \bibinfo{journal}{\emph{arXiv preprint arXiv:2101.12127}}
  (\bibinfo{year}{2021}).
\newblock


\bibitem[Nagrecha(2021)]%
        {nagrecha2021model}
\bibfield{author}{\bibinfo{person}{Kabir Nagrecha}.}
  \bibinfo{year}{2021}\natexlab{}.
\newblock \showarticletitle{Model-parallel model selection for deep learning
  systems}. In \bibinfo{booktitle}{\emph{Proceedings of the 2021 International
  Conference on Management of Data}}. \bibinfo{pages}{2929--2931}.
\newblock


\bibitem[Nagrecha(2023)]%
        {researchExam}
\bibfield{author}{\bibinfo{person}{Kabir Nagrecha}.}
  \bibinfo{year}{2023}\natexlab{}.
\newblock \bibinfo{title}{Systems for Parallel and Distributed Large-Model Deep
  Learning Training}.
\newblock
\newblock


\bibitem[Nagrecha and Kumar(2022)]%
        {nagrecha2022hydra}
\bibfield{author}{\bibinfo{person}{Kabir Nagrecha} {and} \bibinfo{person}{Arun
  Kumar}.} \bibinfo{year}{2022}\natexlab{}.
\newblock \bibinfo{title}{Hydra: A System for Large Multi-Model Deep Learning}.
\newblock
\newblock
\showeprint[arxiv]{2110.08633}~[cs.DC]


\bibitem[Nagrecha and Kumar(2023)]%
        {nagrechasaturn}
\bibfield{author}{\bibinfo{person}{Kabir Nagrecha} {and} \bibinfo{person}{Arun
  Kumar}.} \bibinfo{year}{2023}\natexlab{}.
\newblock \showarticletitle{Saturn: An Optimized Data System for
  Multi-Large-Model Deep Learning Workloads (Information System
  Architectures)}.
\newblock  (\bibinfo{year}{2023}).
\newblock


\bibitem[Nakandala and Kumar(2022)]%
        {nakandala2022nautilus}
\bibfield{author}{\bibinfo{person}{Supun Nakandala} {and} \bibinfo{person}{Arun
  Kumar}.} \bibinfo{year}{2022}\natexlab{}.
\newblock \showarticletitle{Nautilus: An Optimized System for Deep Transfer
  Learning over Evolving Training Datasets}. In
  \bibinfo{booktitle}{\emph{Proceedings of the 2022 International Conference on
  Management of Data}}. \bibinfo{pages}{506--520}.
\newblock


\bibitem[Nakandala et~al\mbox{.}(2020)]%
        {nakandala2020incremental}
\bibfield{author}{\bibinfo{person}{Supun Nakandala}, \bibinfo{person}{Kabir
  Nagrecha}, \bibinfo{person}{Arun Kumar}, {and} \bibinfo{person}{Yannis
  Papakonstantinou}.} \bibinfo{year}{2020}\natexlab{}.
\newblock \showarticletitle{Incremental and approximate computations for
  accelerating deep CNN inference}.
\newblock \bibinfo{journal}{\emph{ACM Transactions on Database Systems (TODS)}}
  \bibinfo{volume}{45}, \bibinfo{number}{4} (\bibinfo{year}{2020}),
  \bibinfo{pages}{1--42}.
\newblock


\bibitem[Narayanan et~al\mbox{.}(2020)]%
        {narayanan2020heterogeneity}
\bibfield{author}{\bibinfo{person}{Deepak Narayanan}, \bibinfo{person}{Keshav
  Santhanam}, \bibinfo{person}{Fiodar Kazhamiaka}, \bibinfo{person}{Amar
  Phanishayee}, {and} \bibinfo{person}{Matei Zaharia}.}
  \bibinfo{year}{2020}\natexlab{}.
\newblock \showarticletitle{Heterogeneity-aware cluster scheduling policies for
  deep learning workloads}. In \bibinfo{booktitle}{\emph{Proceedings of the
  14th USENIX Conference on Operating Systems Design and Implementation}}.
  \bibinfo{pages}{481--498}.
\newblock


\bibitem[Olukotun(2022)]%
        {virtuouscycle}
\bibfield{author}{\bibinfo{person}{Kunle Olukotun}.}
  \bibinfo{year}{2022}\natexlab{}.
\newblock \bibinfo{title}{Systems for ML and ML for Systems: A Virtuous Cycle}.
   (\bibinfo{year}{2022}).
\newblock
\urldef\tempurl%
\url{https://mlsys.org/virtual/2022/invited-talk/2065}
\showURL{%
\tempurl}
\newblock
\shownote{MLSys}.


\bibitem[Ortiz et~al\mbox{.}(2018)]%
        {ortiz2018learning}
\bibfield{author}{\bibinfo{person}{Jennifer Ortiz}, \bibinfo{person}{Magdalena
  Balazinska}, \bibinfo{person}{Johannes Gehrke}, {and}
  \bibinfo{person}{S.~Sathiya Keerthi}.} \bibinfo{year}{2018}\natexlab{}.
\newblock \bibinfo{title}{Learning State Representations for Query Optimization
  with Deep Reinforcement Learning}.
\newblock
\newblock
\urldef\tempurl%
\url{https://doi.org/10.48550/ARXIV.1803.08604}
\showDOI{\tempurl}


\bibitem[Penedo et~al\mbox{.}(2023)]%
        {penedo2023refinedweb}
\bibfield{author}{\bibinfo{person}{Guilherme Penedo}, \bibinfo{person}{Quentin
  Malartic}, \bibinfo{person}{Daniel Hesslow}, \bibinfo{person}{Ruxandra
  Cojocaru}, \bibinfo{person}{Alessandro Cappelli}, \bibinfo{person}{Hamza
  Alobeidli}, \bibinfo{person}{Baptiste Pannier}, \bibinfo{person}{Ebtesam
  Almazrouei}, {and} \bibinfo{person}{Julien Launay}.}
  \bibinfo{year}{2023}\natexlab{}.
\newblock \bibinfo{title}{The RefinedWeb Dataset for Falcon LLM: Outperforming
  Curated Corpora with Web Data, and Web Data Only}.
\newblock
\newblock
\showeprint[arxiv]{2306.01116}~[cs.CL]


\bibitem[Peng et~al\mbox{.}(2018)]%
        {peng2018optimus}
\bibfield{author}{\bibinfo{person}{Yanghua Peng}, \bibinfo{person}{Yixin Bao},
  \bibinfo{person}{Yangrui Chen}, \bibinfo{person}{Chuan Wu}, {and}
  \bibinfo{person}{Chuanxiong Guo}.} \bibinfo{year}{2018}\natexlab{}.
\newblock \showarticletitle{Optimus: an efficient dynamic resource scheduler
  for deep learning clusters}. In \bibinfo{booktitle}{\emph{Proceedings of the
  Thirteenth EuroSys Conference}}. \bibinfo{pages}{1--14}.
\newblock


\bibitem[Peng et~al\mbox{.}(2021a)]%
        {9328612}
\bibfield{author}{\bibinfo{person}{Yanghua Peng}, \bibinfo{person}{Yixin Bao},
  \bibinfo{person}{Yangrui Chen}, \bibinfo{person}{Chuan Wu},
  \bibinfo{person}{Chen Meng}, {and} \bibinfo{person}{Wei Lin}.}
  \bibinfo{year}{2021}\natexlab{a}.
\newblock \showarticletitle{DL2: A Deep Learning-Driven Scheduler for Deep
  Learning Clusters}.
\newblock \bibinfo{journal}{\emph{IEEE Transactions on Parallel and Distributed
  Systems}} \bibinfo{volume}{32}, \bibinfo{number}{8} (\bibinfo{year}{2021}),
  \bibinfo{pages}{1947--1960}.
\newblock
\urldef\tempurl%
\url{https://doi.org/10.1109/TPDS.2021.3052895}
\showDOI{\tempurl}


\bibitem[Peng et~al\mbox{.}(2021b)]%
        {peng2021dl2}
\bibfield{author}{\bibinfo{person}{Yanghua Peng}, \bibinfo{person}{Yixin Bao},
  \bibinfo{person}{Yangrui Chen}, \bibinfo{person}{Chuan Wu},
  \bibinfo{person}{Chen Meng}, {and} \bibinfo{person}{Wei Lin}.}
  \bibinfo{year}{2021}\natexlab{b}.
\newblock \showarticletitle{Dl2: A deep learning-driven scheduler for deep
  learning clusters}.
\newblock \bibinfo{journal}{\emph{IEEE Transactions on Parallel and Distributed
  Systems}} \bibinfo{volume}{32}, \bibinfo{number}{8} (\bibinfo{year}{2021}),
  \bibinfo{pages}{1947--1960}.
\newblock


\bibitem[Qiao et~al\mbox{.}(2021)]%
        {qiao2021pollux}
\bibfield{author}{\bibinfo{person}{Aurick Qiao}, \bibinfo{person}{Sang~Keun
  Choe}, \bibinfo{person}{Suhas~Jayaram Subramanya}, \bibinfo{person}{Willie
  Neiswanger}, \bibinfo{person}{Qirong Ho}, \bibinfo{person}{Hao Zhang},
  \bibinfo{person}{Gregory~R Ganger}, {and} \bibinfo{person}{Eric~P Xing}.}
  \bibinfo{year}{2021}\natexlab{}.
\newblock \showarticletitle{Pollux: Co-adaptive Cluster Scheduling for
  Goodput-Optimized Deep Learning.}. In \bibinfo{booktitle}{\emph{OSDI}},
  Vol.~\bibinfo{volume}{21}. \bibinfo{pages}{1--18}.
\newblock


\bibitem[Ren et~al\mbox{.}(2021)]%
        {zerooffload2021}
\bibfield{author}{\bibinfo{person}{Jie Ren}, \bibinfo{person}{Samyam
  Rajbhandari}, \bibinfo{person}{Reza~Yazdani Aminabadi},
  \bibinfo{person}{Olatunji Ruwase}, \bibinfo{person}{Shuangyan Yang},
  \bibinfo{person}{Minjia Zhang}, \bibinfo{person}{Dong Li}, {and}
  \bibinfo{person}{Yuxiong He}.} \bibinfo{year}{2021}\natexlab{}.
\newblock \bibinfo{title}{ZeRO-Offload: Democratizing Billion-Scale Model
  Training}.
\newblock
\newblock
\urldef\tempurl%
\url{https://doi.org/10.48550/ARXIV.2101.06840}
\showDOI{\tempurl}


\bibitem[Sethi et~al\mbox{.}(2022)]%
        {sethi2022recshard}
\bibfield{author}{\bibinfo{person}{Geet Sethi}, \bibinfo{person}{Bilge Acun},
  \bibinfo{person}{Niket Agarwal}, \bibinfo{person}{Christos Kozyrakis},
  \bibinfo{person}{Caroline Trippel}, {and} \bibinfo{person}{Carole-Jean Wu}.}
  \bibinfo{year}{2022}\natexlab{}.
\newblock \showarticletitle{RecShard: statistical feature-based memory
  optimization for industry-scale neural recommendation}. In
  \bibinfo{booktitle}{\emph{Proceedings of the 27th ACM International
  Conference on Architectural Support for Programming Languages and Operating
  Systems}}. \bibinfo{pages}{344--358}.
\newblock


\bibitem[Shoeybi et~al\mbox{.}(2019)]%
        {megatron2019}
\bibfield{author}{\bibinfo{person}{Mohammad Shoeybi}, \bibinfo{person}{Mostofa
  Patwary}, \bibinfo{person}{Raul Puri}, \bibinfo{person}{Patrick LeGresley},
  \bibinfo{person}{Jared Casper}, {and} \bibinfo{person}{Bryan Catanzaro}.}
  \bibinfo{year}{2019}\natexlab{}.
\newblock \bibinfo{title}{Megatron-LM: Training Multi-Billion Parameter
  Language Models Using Model Parallelism}.
\newblock
\newblock
\urldef\tempurl%
\url{https://doi.org/10.48550/ARXIV.1909.08053}
\showDOI{\tempurl}


\bibitem[Tesauro et~al\mbox{.}(2005)]%
        {Tesauro2005OnlinePM}
\bibfield{author}{\bibinfo{person}{Gerald Tesauro}, \bibinfo{person}{Rajarshi
  Das}, {and} \bibinfo{person}{Nicholas~K. Jong}.}
  \bibinfo{year}{2005}\natexlab{}.
\newblock \showarticletitle{Online Performance Management Using Hybrid
  Reinforcement Learning}.
\newblock


\bibitem[Thrun and Schwartz(1993)]%
        {thrun1993issues}
\bibfield{author}{\bibinfo{person}{Sebastian Thrun} {and}
  \bibinfo{person}{Anton Schwartz}.} \bibinfo{year}{1993}\natexlab{}.
\newblock \showarticletitle{Issues in using function approximation for
  reinforcement learning}. In \bibinfo{booktitle}{\emph{Proceedings of the
  Fourth Connectionist Models Summer School}}, Vol.~\bibinfo{volume}{255}.
  Hillsdale, NJ, \bibinfo{pages}{263}.
\newblock


\bibitem[Wang et~al\mbox{.}(2021a)]%
        {wang2021sensai}
\bibfield{author}{\bibinfo{person}{Guanhua Wang}, \bibinfo{person}{Zhuang Liu},
  \bibinfo{person}{Brandon Hsieh}, \bibinfo{person}{Siyuan Zhuang},
  \bibinfo{person}{Joseph Gonzalez}, \bibinfo{person}{Trevor Darrell}, {and}
  \bibinfo{person}{Ion Stoica}.} \bibinfo{year}{2021}\natexlab{a}.
\newblock \showarticletitle{sensai: Convnets decomposition via class
  parallelism for fast inference on live data}.
\newblock \bibinfo{journal}{\emph{Proceedings of Machine Learning and Systems}}
   \bibinfo{volume}{3} (\bibinfo{year}{2021}), \bibinfo{pages}{664--679}.
\newblock


\bibitem[Wang et~al\mbox{.}(2021b)]%
        {wang2021gradient}
\bibfield{author}{\bibinfo{person}{Pei Wang}, \bibinfo{person}{Kabir Nagrecha},
  {and} \bibinfo{person}{Nuno Vasconcelos}.} \bibinfo{year}{2021}\natexlab{b}.
\newblock \showarticletitle{Gradient-based algorithms for machine teaching}. In
  \bibinfo{booktitle}{\emph{Proceedings of the IEEE/CVF Conference on Computer
  Vision and Pattern Recognition}}. \bibinfo{pages}{1387--1396}.
\newblock


\bibitem[Wang et~al\mbox{.}(2022)]%
        {wang2022merlin}
\bibfield{author}{\bibinfo{person}{Zehuan Wang}, \bibinfo{person}{Yingcan Wei},
  \bibinfo{person}{Minseok Lee}, \bibinfo{person}{Matthias Langer},
  \bibinfo{person}{Fan Yu}, \bibinfo{person}{Jie Liu}, \bibinfo{person}{Shijie
  Liu}, \bibinfo{person}{Daniel~G Abel}, \bibinfo{person}{Xu Guo},
  \bibinfo{person}{Jianbing Dong}, {et~al\mbox{.}}}
  \bibinfo{year}{2022}\natexlab{}.
\newblock \showarticletitle{Merlin hugectr: Gpu-accelerated recommender system
  training and inference}. In \bibinfo{booktitle}{\emph{Proceedings of the 16th
  ACM Conference on Recommender Systems}}. \bibinfo{pages}{534--537}.
\newblock


\bibitem[Wei et~al\mbox{.}(2022)]%
        {Wei_2022}
\bibfield{author}{\bibinfo{person}{Yingcan Wei}, \bibinfo{person}{Matthias
  Langer}, \bibinfo{person}{Fan Yu}, \bibinfo{person}{Minseok Lee},
  \bibinfo{person}{Jie Liu}, \bibinfo{person}{Ji Shi}, {and}
  \bibinfo{person}{Zehuan Wang}.} \bibinfo{year}{2022}\natexlab{}.
\newblock \showarticletitle{A {GPU}-specialized Inference Parameter Server for
  Large-Scale Deep Recommendation Models}. In
  \bibinfo{booktitle}{\emph{Sixteenth {ACM} Conference on Recommender
  Systems}}. \bibinfo{publisher}{{ACM}}.
\newblock
\urldef\tempurl%
\url{https://doi.org/10.1145/3523227.3546765}
\showDOI{\tempurl}


\bibitem[Wu et~al\mbox{.}(2021)]%
        {wu2021sustainable}
\bibfield{author}{\bibinfo{person}{Carole-Jean Wu}, \bibinfo{person}{Ramya
  Raghavendra}, \bibinfo{person}{Udit Gupta}, \bibinfo{person}{Bilge Acun},
  \bibinfo{person}{Newsha Ardalani}, \bibinfo{person}{Kiwan Maeng},
  \bibinfo{person}{Gloria Chang}, \bibinfo{person}{Fiona~Aga Behram},
  \bibinfo{person}{James Huang}, \bibinfo{person}{Charles Bai},
  \bibinfo{person}{Michael Gschwind}, \bibinfo{person}{Anurag Gupta},
  \bibinfo{person}{Myle Ott}, \bibinfo{person}{Anastasia Melnikov},
  \bibinfo{person}{Salvatore Candido}, \bibinfo{person}{David Brooks},
  \bibinfo{person}{Geeta Chauhan}, \bibinfo{person}{Benjamin Lee},
  \bibinfo{person}{Hsien-Hsin~S. Lee}, \bibinfo{person}{Bugra Akyildiz},
  \bibinfo{person}{Maximilian Balandat}, \bibinfo{person}{Joe Spisak},
  \bibinfo{person}{Ravi Jain}, \bibinfo{person}{Mike Rabbat}, {and}
  \bibinfo{person}{Kim Hazelwood}.} \bibinfo{year}{2021}\natexlab{}.
\newblock \bibinfo{title}{Sustainable AI: Environmental Implications,
  Challenges and Opportunities}.
\newblock
\newblock
\urldef\tempurl%
\url{https://doi.org/10.48550/ARXIV.2111.00364}
\showDOI{\tempurl}


\bibitem[Xiao et~al\mbox{.}(2018)]%
        {xiao2018gandiva}
\bibfield{author}{\bibinfo{person}{Wencong Xiao}, \bibinfo{person}{Romil
  Bhardwaj}, \bibinfo{person}{Ramachandran Ramjee}, \bibinfo{person}{Muthian
  Sivathanu}, \bibinfo{person}{Nipun Kwatra}, \bibinfo{person}{Zhenhua Han},
  \bibinfo{person}{Pratyush Patel}, \bibinfo{person}{Xuan Peng},
  \bibinfo{person}{Hanyu Zhao}, \bibinfo{person}{Quanlu Zhang},
  {et~al\mbox{.}}} \bibinfo{year}{2018}\natexlab{}.
\newblock \showarticletitle{Gandiva: Introspective cluster scheduling for deep
  learning}. In \bibinfo{booktitle}{\emph{13th USENIX Symposium on Operating
  Systems Design and Implementation (OSDI 18)}}. \bibinfo{pages}{595--610}.
\newblock


\bibitem[Xiao et~al\mbox{.}(2020)]%
        {xiao2020antman}
\bibfield{author}{\bibinfo{person}{Wencong Xiao}, \bibinfo{person}{Shiru Ren},
  \bibinfo{person}{Yong Li}, \bibinfo{person}{Yang Zhang},
  \bibinfo{person}{Pengyang Hou}, \bibinfo{person}{Zhi Li},
  \bibinfo{person}{Yihui Feng}, \bibinfo{person}{Wei Lin}, {and}
  \bibinfo{person}{Yangqing Jia}.} \bibinfo{year}{2020}\natexlab{}.
\newblock \showarticletitle{$\{$AntMan$\}$: Dynamic Scaling on $\{$GPU$\}$
  Clusters for Deep Learning}. In \bibinfo{booktitle}{\emph{14th USENIX
  Symposium on Operating Systems Design and Implementation (OSDI 20)}}.
  \bibinfo{pages}{533--548}.
\newblock


\bibitem[Zhao et~al\mbox{.}(2022)]%
        {Zhao_2022}
\bibfield{author}{\bibinfo{person}{Mark Zhao}, \bibinfo{person}{Niket Agarwal},
  \bibinfo{person}{Aarti Basant}, \bibinfo{person}{Bu{\u{g}}ra Gedik},
  \bibinfo{person}{Satadru Pan}, \bibinfo{person}{Mustafa Ozdal},
  \bibinfo{person}{Rakesh Komuravelli}, \bibinfo{person}{Jerry Pan},
  \bibinfo{person}{Tianshu Bao}, \bibinfo{person}{Haowei Lu},
  \bibinfo{person}{Sundaram Narayanan}, \bibinfo{person}{Jack Langman},
  \bibinfo{person}{Kevin Wilfong}, \bibinfo{person}{Harsha Rastogi},
  \bibinfo{person}{Carole-Jean Wu}, \bibinfo{person}{Christos Kozyrakis}, {and}
  \bibinfo{person}{Parik Pol}.} \bibinfo{year}{2022}\natexlab{}.
\newblock \showarticletitle{Understanding data storage and ingestion for
  large-scale deep recommendation model training}. In
  \bibinfo{booktitle}{\emph{Proceedings of the 49th Annual International
  Symposium on Computer Architecture}}. \bibinfo{publisher}{{ACM}}.
\newblock
\urldef\tempurl%
\url{https://doi.org/10.1145/3470496.3533044}
\showDOI{\tempurl}


\bibitem[Zhong et~al\mbox{.}(2017)]%
        {zhong2017seq}
\bibfield{author}{\bibinfo{person}{Victor Zhong}, \bibinfo{person}{Caiming
  Xiong}, {and} \bibinfo{person}{Richard Socher}.}
  \bibinfo{year}{2017}\natexlab{}.
\newblock \bibinfo{title}{Seq2SQL: Generating Structured Queries from Natural
  Language using Reinforcement Learning}.
\newblock
\newblock
\urldef\tempurl%
\url{https://doi.org/10.48550/ARXIV.1709.00103}
\showDOI{\tempurl}


\end{thebibliography}

\end{document}